\documentclass[12pt, draftclsnofoot, onecolumn]{IEEEtran}
\ifCLASSINFOpdf
\else
\fi

\usepackage{cite}
\usepackage{stfloats}
\ifCLASSINFOpdf
   \usepackage[pdftex]{graphicx}
   \graphicspath{{../pdf/}{../jpeg/}}
   \DeclareGraphicsExtensions{.pdf,.jpeg,.png}
\else
   \usepackage[dvips]{graphicx}
   \graphicspath{{../eps/}}
   \DeclareGraphicsExtensions{.eps}
\fi

\usepackage[cmex10]{amsmath}
\usepackage{mathrsfs,color,amssymb}
\usepackage{amsfonts}

\usepackage{algorithm}
\usepackage{algorithmic}
\allowdisplaybreaks

\ifCLASSOPTIONcompsoc
\else
  \usepackage[caption=false,font=footnotesize]{subfig}
\fi

\usepackage{hyperref}
\usepackage{graphicx}
\usepackage{epstopdf}

\newtheorem{theorem}{\bf{Theorem}}
\newtheorem{lemma}{\bf{Lemma}}
\newtheorem{corollary}{\bf{Corollary}}

\newtheorem{remark}{\textit{Remark}}

\begin{document}

%

\title{Performance Analysis for Massive MIMO Downlink with Low Complexity Approximate Zero-Forcing Precoding}
%
%
%

\author{Cheng~Zhang,~\IEEEmembership{Student Member,~IEEE,}
        Yindi~Jing,~\IEEEmembership{Member,~IEEE,}
        Yongming~Huang,~\IEEEmembership{Member,~IEEE,} and Luxi Yang,~\IEEEmembership{Member,~IEEE}
\thanks{
This work has been submitted to the IEEE for possible publication. Copyright may be transferred without notice, after which this version may no longer be accessible.

Part of this work has been submitted to \textit{IEEE International Conference on Communications} (ICC) 2018.

C. Zhang, Y. Huang, and L. Yang are with the National Mobile Communications Research Laboratory, School of Information Science and Engineering, Southeast University, Nanjing 210096, P. R. China (email: {zhangcheng1988, huangym, lxyang}@seu.edu.cn).

Y. Jing is with the Department of Electrical and Computer Engineering, University of Alberta, Edmonton, Canada, T6G 1H9 (email: yindi@ualberta.ca).
}}

\maketitle
\vspace{-1.3cm}
\begin{abstract}
Zero-forcing (ZF) precoding plays an important role for massive MIMO downlink due to its near optimal performance. However, the high computation cost of the involved matrix inversion hinders its application.
In this paper, we adopt the first order Neumann series (NS) for a low-complexity approximation.
By introducing a relaxation parameter jointly with one selected user's interference to others into the precondition matrix, we propose the identity-plus-column NS (ICNS) method. By further exploiting the multi-user diversity gain via choosing the user with the largest interference to others, the ordered ICNS method is also proposed.
Moreover, the sum-rate approximations of the proposed ICNS method and the competitive existing identity matrix based NS (INS) method are derived in closed-form, based on which the performance loss of ICNS due to inversion approximation compared with ideal ZF and its performance gain over INS are explicitly analyzed for three typical massive MIMO scenarios. Finally, simulations verify our analytical results and also show that the proposed two designs achieve better performance-complexity tradeoff than ideal ZF and existing low-complexity ZF precodings for practical large antenna number, correlated channels and not-so-small loading factor.
\end{abstract}

\vspace{-0.5cm}

%
\IEEEpeerreviewmaketitle

\section{Introduction}
As a promising key technology for future cellular network, massive multiple-input multiple-output (MIMO) has been widely studied in recent years \cite{Marzetta_non, Rusek_Scaling, Emil_Massive}. By deploying large-scale antenna array at the base station (BS), great increase in array gain and spatial resolution can be achieved which results in higher spectrum efficiency and the capability of serving more users simultaneously \cite{Zhang_sum,Cheng_Performance,Lu_low}. Existing works \cite{Rusek_Scaling}, \cite{Prabhu_Approx} show that linear precoding techniques such as zero-forcing (ZF) can achieve the performance of the capacity-approaching schemes, e.g., the dirty paper coding or other advanced non-linear precoding methods, in the favorable channel condition, i.e., users have asymptotic orthogonal channels as the number of BS antennas grows large. ZF precoding has higher computational efficiency than its non-linear alternatives. However, since it involves the inverse of the Gram matrix of all users' channel vectors, the number of multiplication and division operations are cubic and quadratic in the number of users \cite{Prabhu_Approx,Wu_large}, if conventional inversion methods are used, e.g., via orthogonal and upper triangular (QR) decomposition using Gram-Schmidt process or Givens rotation, and Gauss-Jordan elimination \cite{Prabhu_Approx}.
In massive MIMO systems,  the user number tends to be large, making the computational complexity of ZF precoding prohibitive.

Recently, many efforts have been endeavored to further reduce the complexity of ZF precoding. The first class of methods uses the Neumann series (NS) expansion to transform the inverse of
the Gram matrix into that of a simple precondition matrix and some simple matrix multiplications and summations.
Two designs were studied in \cite{Prabhu_Approx}, where the precondition matrix is set to be a scaled identity matrix (referred to as the INS design) and the diagonal matrix made up by the main diagonal of the Gram matrix (referred to as the DNS design). The DNS design was shown to have better performance than INS when the Gram matrix is strongly diagonal dominant. However, when the diagonal dominance of the Gram matrix is not strong due to either high channel correlation or limited number of BS antennas, DNS causes large performance degradation \cite{Prabhu_Hardware}.
To solve this problem, a tri-digonal precondition matrix was proposed in \cite{Prabhu_Hardware} by adding secondary diagonal lines of the Gram matrix (referred to as the TNS design). However, the complexity of the inversion of tri-diagonal precondition matrix itself becomes a problem \cite{Abbas_Low}. Consequently, a new design was proposed in \cite{Abbas_Low} where the precondition matrix is formed by the non-diagonal elements of the first column of the Gram matrix in addition to its diagonal elements (referred to as the CNS design). Although CNS simplifies the inverse of the precondition matrix, it also has non-negligible performance degradation compared with TNS. Therefore, there is still room to improve for a better balance between the computation complexity and the precoding performance.

The second class of methods uses numerical iterative schemes for solving linear equations \cite{Björck_Numerical}. Instead of first computing the inverse approximation of the Gram matrix and then multiplying it with the symbol vector to obtain the precoded vector, this kind of methods takes the symbol vector as the input and output the precoded vector via certain number of iterations. Typical iterative schemes include Richardson method \cite{Lu_Richardson}, Jacobi method \cite{Song_Joint}, Gauss-Seidel method \cite{Gao_Capacity}, successive over relaxation method \cite{Xie_SOR}, and symmetric successive over relaxation method \cite{Xie_low}. However, while its computational load is advantageous for fast-fading systems, the class of methods has prohibitive computation overhead for systems with moderate to large channel coherence time \cite{Xie_low}, especially for systems with large bandwidth.

Besides the low-complexity ZF precoding design itself, the related analytical performance analysis is also important in the sense of both quantitatively understanding the performance loss due to the inversion approximation and providing explicit expression for parameter optimization. However, few results on performance analysis were provided in existing works.
\cite{Zhu_inversion} studied the effect of the loading factor on both the asymptotic convergence speed of the NS expansion with the DNS precondition matrix design and the mean square error (MSE) between the noiseless received signals with the ideal ZF precoding and that with the DNS procoding.
In \cite{Muller_TPE}, a low-complexity regularized ZF (RZF) precoding was proposed in which the matrix inversion is replaced by a truncated polynomial expansion (TPE). An asymptotic deterministic expression of the signal-to-interference-plus-noise ratio (SINR) was derived using random matrix theory. Meanwhile, a closed-form expression was given for the polynomial coefficients that maximizes this SINR expression.

In this paper, we consider the first kind of NS based low-complexity design for practical scenarios with not-so-small loading factor and/or high channel correlation, where existing designs suffer considerable performance degradation. Further, different from most existing works, we focus on the case of the first-order NS. This is because when the order number of NS is larger than one, the computational complexity is comparable to that of conventional inversion methods \cite{Prabhu_Approx, Wang_efficient}.
Specifically, by observing the good performance-complexity tradeoff of the CNS method and the strong robustness of the INS method, we first propose the identity-plus-column NS (ICNS) method by replacing the diagonal elements of CNS's precondition matrix with a relaxation parameter. Then a channel-correlation-adaptive design for the relaxation parameter is given. Both the relaxation parameter and the non-diagonal elements of the first column of the precondition matrix can help to handle the effect of user interference on the inversion approximation more carefully. Further, by choosing the user with the largest interference to others, the ordered ICNS method is proposed to exploit the multi-user diversity.

Further, we provide comprehensive performance analysis on the sum-rate directly, while existing performance studies were on the inversion approximation error. A sum-rate approximation of the proposed ICNS scheme is derived in closed-form for the correlated massive MIMO channel. In addition, we provide a closed-form sum-rate approximation for the most competitive benchmark, the INS scheme. And our analytical method also applies to
other existing low-complexity ZF precodings.
Based on these analytical results, the comparison between the proposed ICNS scheme, the INS scheme, the ideal ZF and maximal ratio transmission (MRT) are elaborated for three typical cases in massive MIMO systems, i.e., 1) asymptotically large BS antenna number and user number with fixed ratio; 2) finite user number and large but finite BS antenna number, and 3) finite user number and asymptotically large BS antenna number.

Comparison results show that
1) for Case 1, ICNS outperforms INS with intermediate loading factor, while with either low or high loading factor the advantage becomes negligible.
Meanwhile, the favorable range of loading factor for ICNS to have comparable sum-rate to the ideal ZF is derived in closed-form.
2) For Case 2, the sum-rate of ICNS is better than that of INS and the advantage first increases with BS antenna number and then decreases to zero as BS antenna number further grows.
3) For Case 3, the sum-rates of ICNS and INS both approach that of the ideal ZF. However, the sum-rate of MRT has much slower convergence rate compared with the above three schemes.
Simulation results validate the derived sum-rate approximations and the analytical comparison between ICNS and INS. Meanwhile, with the help of the complexity analysis, it is shown that the proposed ICNS and ordered ICNS can achieve better complexity-performance tradeoff compared with existing low-complexity ZF precodings for massive MIMO systems with correlated channels, practical antenna number and not-so-small loading factor.

The remaining of the paper is organized as follows. In Section II, the system model is introduced along with the low-complexity ZF precoding problem and existing designs. Section III gives the proposed low-complexity approximate designs, i.e., the ICNS and ordered ICNS methods, and their computational complexity analysis. In Section IV, closed-form sum-rate approximations are derived for both INS and ICNS based on which a comprehensive performance comparison is provided. Section V shows simulations and conclusions are given in Section VI.

In this paper, bold upper case letters and bold lower case letters are used to denote matrices and vectors, respectively. For a matrix $\bf A$, its conjugate transpose, transpose, and trace are denoted by ${\bf A}^H$, ${\bf A}^T$ and ${\rm tr}\{{\bf A}\}$, respectively. $[{\bf A}]_{i,j}$ is the $(i,j)$th entry of $\bf A$. ${\bf I}_M$ denotes the $M$ dimensional identity matrix. $\|{\bf A}\|_F$ denotes the Frobenius norm of $\bf A$. ${\mathcal {CN}}({\bf 0},{\bf \Sigma})$ denotes the circularly symmetric complex Gaussian distribution with mean $\bf 0$ and covariance matrix $\bf \Sigma$. ${\rm E}\{\cdot\}$ is the mean operator. ${a} = \mathcal{O}\left( {{b}} \right)$ means that $a$ and $b$ have the same scaling with respect to an asymptotic parameter given in the context. $\lambda_{max}(\bf \cdot)$ denotes the spectral norm operator.

\section{System Model and Problem Statement}
\subsection{System Model}
We consider a single-cell downlink system where a BS, equipped with $M$ antennas, serves $K$ single-antenna users and $M \ge K \gg 1$. Let $r=K/M$, which is the loading factor. Let ${\bf h}_k^H$ be the downlink channel from the BS to User $k$ which can be written as
\begin{equation}\label{channel-model}
{\bf h}_k={\bf R}^{1/2}{\bf z}_k,
\end{equation}
where  ${\bf z}_k\sim \mathcal{CN}({\bf 0},{\bf I}_M)$ is the fast-fading channel vector and ${\bf R}\in \mathbb{C}^{M\times M}$ denotes the channel covariance matrix with large scale fading normalization ${\rm tr}\{{\bf R}\}=M$. Specifically,
${\bf R}$ is modeled as in \cite{Hoydis_Massive}:
\begin{equation}\label{correlation model}
{\bf{R}} = \frac{1}{c}{\bf{A}}{\bf{A}}^H,
\end{equation}
where the channel direction matrix ${\bf A}$ is an ${M \times c{M}}$ semi-unitary matrix and $c\in(0,1]$ indicates the channel correlation level. For example, elements of the channel vector become independent and identically distributed (i.i.d.) when $c = 1$. With the models in \eqref{channel-model} and \eqref{correlation model}, all users' channel covariances are assumed to be the same, and the power beam spectrum (PBS) is assumed to be flat along the effective channel directions. The former assumption is applicable when the antenna correlation is mainly dependent on the BS inter-element antenna spacing as in the exponential model \cite{Clerckx_correlated} or the local scatterers at the BS rather than those at the users \cite{Rao_distributed}. The motivation for the latter assumption is two-fold \cite{Zhang_sum}. First, while the PBS in general can have many possible profiles in practice, the flat PBS model can serve as an approximation of the average effect of all possible profiles. Secondly, as explained in \cite[Sec. IV]{Hoydis_Massive}, when the antenna aperture increases with each additional antenna element and $c$ depends on the amount of scattering in the channel, this model is applicable. Define the channel matrix as ${\bf H}=[{\bf h}_1,...,{\bf h}_K]$ where channel vectors of different users are assumed to be independent.

The received signal $y_k$ at User $k$ is given by
\begin{equation}\label{eq1}
y_k=\sqrt{\rho_t}{\bf h}_k^H{\bf W}{\bf s}+ n_k, \hspace{10pt} k=1,...,K,
\end{equation}
where $\rho_t$ is the average transmit power, $n_k$'s are i.i.d. noises each following ${\mathcal{CN}}(0,1)$, ${\bf W}=[{\bf w}_1,...,{\bf w}_K]\in \mathbb{C}^{M\times K}$ is the precoding matrix, and ${\bf s}=[s_1,...,s_K]^T\sim \mathcal{CN}(0,{\bf I}_K)$ is the vector containing all users' symbols. The precoding matrix is normalized as
\begin{equation}\label{eq3_1}
{\rm E}\{{\rm tr}\{{\bf W}{\bf W}^H\}\}=1.
\end{equation}

Consequently, the SINR at User $k$ is
\begin{equation}\label{eq3}
{\rm SINR}_k=\frac{{\bf h}_k^H{\bf w}_k{\bf w}_k^H{\bf h}_k}{{\bf h}_k^H{\bf W}_k{\bf W}_k^H{\bf h}_k+1/{\rho_t}},
\end{equation}
where ${\bf W}_k=[{\bf w}_1,...,{\bf w}_{k-1},{\bf w}_{k+1},...,{\bf w}_K]$.


\subsection{The Low Complexity Precoding Design Problem}\label{problem}
The ZF precoding can be represented as
\begin{equation}\label{perfect ZF}
{\bf W}_{ZF}=\beta_{ZF}{\bf H}({\bf H}^H{\bf H})^{-1},
\end{equation}
where the power normalization parameter $\beta_{ZF}$ is set such that ${\bf W}_{ZF}$ satisfies the power constraint in \eqref{eq3_1}. A disadvantage of ZF precoding is its high computational load, mainly caused by the matrix inversion. For conventional QR decomposition based methods, the matrix inversion has the complexity of ${\mathcal O}(K^3)$ complex multiplications and ${\mathcal O}(K^2)$ complex divisions \cite{Prabhu_Approx}, which can be prohibitive for massive MIMO with large $K$. \textit{Our main goal} is to find an appropriate approximation of the matrix inversion with low computational complexity.

Define the Gram matrix ${\bf G}={\bf H}^H{\bf H}/{M}$. The inverse of ${\bf G}$ can be expressed as its NS:
\begin{equation}\label{Neumann series expansion}
  {\bf G}^{-1}=\sum_{n=0}^{\infty}(-{\bf D}^{-1}{\bf E})^n{\bf D}^{-1}
\end{equation}
if the precondition matrix ${\bf D}$ satisfies
\begin{equation}\label{spectrum-norm}
\lim_{n\rightarrow\infty}\left(-{\bf D}^{-1}{\bf E}\right)^n={\bf 0} \quad \text{or} \quad \lambda_{max}(-{\bf D}^{-1}{\bf E})<1,
\end{equation}
where ${\bf E}={\bf G}-{\bf D}$. Thus, a natural approximation of ${\bf G}^{-1}$ is
\begin{equation}\label{approx_NS}
{\bf G}^{-1}\approx\sum_{n=0}^{L}(-{\bf D}^{-1}{\bf E})^n{\bf D}^{-1},
\end{equation}
where the approximation becomes equality when the order number $L$ grows to infinity.
The calculations of the approximation in \eqref{approx_NS} involve the inversion of $\bf D$ and some matrix multiplications and summations. Larger $L$ means better approximation performance but at the same time higher complexity. Notice that for $L>1$, the multiplication of $K\times K$ dimensional square matrices is unavoidable, making the computational complexity of the NS based approximation comparable to that of conventional inverse methods. Thus, we consider the case of $L=1$ only. Corresponding, the approximate precoding matrix is
\begin{equation}\label{general_lowP}
{\bf W}_{ZF}\hspace{-0.1cm}\approx\hspace{-0.1cm}{\bf W}\hspace{-0.1cm}=\hspace{-0.1cm}\frac{\beta}{M}{\bf H}\hspace{-0.1cm}\left({\bf D}^{-1}\hspace{-0.1cm}-\hspace{-0.1cm}{\bf D}^{-1}{\bf E}{\bf D}^{-1}\hspace{-0.05cm}\right)\hspace{-0.1cm}
=\hspace{-0.1cm} \frac{{{\beta}}}{M}{\bf{H}}\hspace{-0.1cm}\left( {2{{\bf{D}}^{ - 1}} \hspace{-0.1cm}-\hspace{-0.1cm} {{\bf{D}}^{ - 1}}{\bf{G}}{{\bf{D}}^{ - 1}}} \hspace{-0.05cm}\right),
\end{equation}
where $\beta$ is set such that ${\bf W}$ satisfies the power constraint in \eqref{eq3_1}.
The choice of the precondition matrix $\bf D$ is critical for the performance-complexity tradeoff of this approximation. A complex structure for $\bf D$ may improve the approximation performance, but the corresponding computational complexity becomes a problem.

\subsection{Existing Designs for the Precondition Matrix}
Several typical existing designs are introduced as follows.
\subsubsection{\textbf{INS} Method} The INS method has the following precondition matrix
\begin{equation}\label{D_INS}
{\bf D}_{I}=\omega_I {\bf I}_K.
\end{equation}
To maximize the asymptotic convergence speed, i.e., minimizing $\lambda_{max}(-{\bf D}_I^{-1}{\bf E}_I)$,
the relaxation parameter $\omega_I$ can be given as \cite{Björck_Numerical}
\begin{equation}\label{optimal-omega-iid}
\omega^{\star}_{I}=\frac{b+a}{2},
\end{equation}
where $a$ and $b$ are the smallest and largest eigenvalue of $\bf G$, respectively. Since the calculation of $a$ and $b$ based on instantaneous $\bf G$ also brings huge computation cost, it is more practical to use the asymptotic value for large $M$. For i.i.d. channels, when $M,K\rightarrow\infty$ with $r=K/M$ being fixed, the asymptotic value of $a$ and $b$ are \cite{Prabhu_Approx}
\begin{equation}\label{a_b_iid}
a=\left(1-\sqrt{r}\right)^2;   b=\left(1+\sqrt{r}\right)^2.
\end{equation}
This asymptotic value for the relaxation parameter was also shown to be effective for asymptotically large $M$ with finite $K$ \cite{Lu_Richardson}.

\subsubsection{\textbf{DNS} Method}
The DNS method has the following diagonal precondition matrix
\begin{equation}
{\bf D}_{D}={\rm diag}_0({\bf G}),
\end{equation}
where
\begin{equation}
[{\rm diag}_n({\bf G})]_{i,j}=\left\{ {\begin{array}{*{20}{c}}
{[{\bf G}]_{i,j}\quad \text{if }|i-j|=n,}\\
{\hspace{-0.5cm}0 \quad \hspace{0.7cm} \text{otherwise.}}
\end{array}} \right.
\end{equation}


\subsubsection{\textbf{TNS} Method}
Via choosing the super diagonal elements and the sub-diagonal elements along with the main diagonal elements of $\bf G$, the precondition matrix of the TNS method is
\begin{equation}
{\bf D}_{T}={\rm diag}_0({\bf G})+ {\rm diag}_1({\bf G}).
\end{equation}

\subsubsection{\textbf{CNS} Method}
For the CNS method, its precondition matrix is composed of the diagonal elements of $\bf G$ and the non-diagonal elements of the $1$st column of $\bf G$, i.e.,
\begin{equation}\label{D_CNS}
{\bf D}_{C}={\rm diag}_0({\bf G})+{\bf G}_c,
\end{equation}
where
\begin{equation}\label{eq18}
[{\bf G}_c]_{i,j}=\left\{ {\begin{array}{*{20}{c}}
{[{\bf G}]_{i,j}\quad \text{if } i>1,j=1,}\\
{\hspace{-0.8cm}0 \quad \hspace{0.7cm} \text{otherwise.}}
\end{array}} \right.
\end{equation}

\section{Proposed Low Complexity Approximate Desgin of ZF Precoding}
By drawing lessons from existing methods, we propose a scheme, called identity-plus-column NS (ICNS) method. Specifically, unlike keeping the diagonal elements of $\bf G$ in CNS, we replace them with a relaxation value $\omega$. Further, the interference from a certain user (denoted as User C) to others are also considered into the construction of the precondition matrix. For ICNS, User C is randomly selected which is equivalent to selecting User $1$ due to the homogeneous channel distribution. The precondition matrix in ICNS can be written as
\begin{equation}\label{proposed precondition}
{\bf D}_A=\omega{\bf I}_K+ {\bf {G}}_c.
\end{equation}
A crucial issue is the design of the relaxation parameter $\omega$. The optimal $\omega$ is the solution for the sum-rate maximization problem. However, the optimization problem is highly challenging due to the difficulty in the sum-rate analysis and the complexity of the sum-rate expression. Instead, a simple heuristic design is to use the asymptotic relaxation parameter for INS.
Since ICNS is equivalent to adding $K-1$ more elements of the $K\times K$ matrix $\bf G$ into the precondition matrix of INS, which is a small change when $K\gg1$, the asymptotic relaxation parameter for INS is expected to have near-optimal performance for ICNS. While the asymptotic relaxation parameter of INS for correlated channels is not available in existing work, we derive it in Lemma \ref{Lemma 1}.
\begin{lemma}\label{Lemma 1}
For the massive MIMO channel with correlation level $c$, when $M,K\rightarrow\infty$ with $r=K/M$ being fixed,
the relaxation parameter for INS that maximizes the convergence speed of the NS is
\begin{equation}\label{heuristic_omega}
\omega^{\star}=({\bar{b}+\bar{a}})/{2},
\end{equation}
where
\begin{equation}\label{bar_a_b}
\bar{a}=\left(1-{\sqrt{\frac{r}{c}}}\right)^2;   \bar{b}=\left(1+{\sqrt{\frac{r}{c}}}\right)^2.
\end{equation}
\end{lemma}
\begin{IEEEproof}
See Appendix A. 
\end{IEEEproof}

Simulation results in Fig. \ref{omega_M60_2} of Section \ref{sr_icns} show that $\omega^{\star}$ has nearly the same performance as the optimal $\omega$ that maximizes the sum-rate.


As explained in Section \ref{problem}, we focus on the practical case of $L=1$ for complexity consideration.
Correspondingly, the precoding matrix of the ICNS method is
\begin{eqnarray}\label{precoding}
\hspace{-0.1cm}{\bf W}_A\hspace{-0.3cm}&=&\hspace{-0.3cm}\frac{\beta_A}{M}{\bf H}\hspace{-0.1cm}\left({\bf D}_A^{-1}\hspace{-0.1cm}-\hspace{-0.1cm}{\bf D}_A^{-1}{\bf E}_A{\bf D}_A^{-1}\hspace{-0.05cm}\right)\hspace{-0.1cm}=\hspace{-0.1cm} \frac{{{\beta_A}}}{M}{\bf{H}}\hspace{-0.1cm}\left( {2{{\bf{D}}_A^{ - 1}} \hspace{-0.1cm}-\hspace{-0.1cm} {{\bf{D}}_A^{ - 1}}{\bf{G}}{{\bf{D}}_A^{ - 1}}} \hspace{-0.05cm}\right).
\end{eqnarray}
where $\beta_A$ is the power normalization parameter for ${{\bf{W}}_A}$ according to \eqref{eq3_1}.

\subsection{Computational Complexity Analysis}\label{comp_ana}
The computational complexity of the proposed ICNS scheme is elaborated as follows. The complexity of the comparison benchmarks, i.e., INS, CNS and TNS, is also provided. Since the multiplication between $\bf H$ and ${\bf D}^{-1}-{\bf D}^{-1}{\bf E}{\bf D}^{-1}$ as shown in \eqref{general_lowP} and the calculation of $\beta$ are common for all methods, we focus on the calculation of ${\bf D}^{-1}-{\bf D}^{-1}{\bf E}{\bf D}^{-1}$ only. Meanwhile, the numbers of multiplication and division operations are used to quantize the computational complexity due to their dominance in computation. Since $K\gg1$,
only the highest order terms of $K$ are kept in the following analysis.

The calculation of ${\bf D}^{-1}-{\bf D}^{-1}{\bf E}{\bf D}^{-1}$ can be divided into two parts, i.e., the calculation of ${\bf D}^{-1}$ and that of ${\bf D}^{-1}{\bf E}{\bf D}^{-1}$.
The first part is studied as follows.
From \eqref{proposed precondition}, we have
\begin{equation}\label{inverse-precond}
{\bf D}_A^{-1}=\frac{1}{\omega}{\bf I}_K - \frac{1}{\omega^2}{\bf {G}}_c.
\end{equation}
Since $w$ can be pre-calculated, i.e., $({1}/{\omega}){\bf I}_K$ is known, the calculation of \eqref{inverse-precond} needs $K$ complex multiplications. Comparatively, since the precondition matrix for CNS in \eqref{D_CNS} can be rewritten as \cite{Abbas_Low}
\begin{equation}\label{D_CNS_rewritten}
{\bf D}_{C}={\rm diag}_0({\bf G})+{\bf G}_c={\rm diag}_0({\bf G})({\bf I}_K+{\bf \tilde G}_c),
\end{equation}
where
\begin{equation}\label{eq18}
[{\bf \tilde G}_c]_{i,j}=\left\{ {\begin{array}{*{20}{c}}
{[{\bf G}]_{i,i}^{-1}[{\bf G}_c]_{i,j}\quad \text{if } i>1,j=1,}\\
{\hspace{-0.5cm}0 \qquad \hspace{1.4cm}\text{otherwise.}}
\end{array}} \right.
\end{equation}
Thus,
\begin{equation}\label{D_CNS_inv}
{\bf D}_{C}^{-1}=({\bf I}_K-{\bf \tilde G}_c)({\rm diag}_0({\bf G}))^{-1}
\end{equation}
and the calculation of ${\bf D}_{C}^{-1}$ needs $K$ complex multiplications and $K$ complex divisions. For INS, since
\begin{equation}\label{D_INS_inv}
{\bf D}_{I}^{-1}=\frac{1}{\omega_I} {\bf I}_K,
\end{equation}
no computation is needed.

For the second part, from \eqref{inverse-precond}, calculating ${\bf D}_A^{-1}{\bf E}_A$ with given ${\bf D}_A^{-1}$ needs $2K^2$ complex multiplications. Then multiplying ${\bf D}_A^{-1}{\bf E}_A$ with ${\bf D}_A^{-1}$ needs another $2K^2$ complex multiplications. All together, ICNS needs $4K^2$ complex multiplications for the second part. For the CNS method, due to the similar structure of ${\bf D}_C^{-1}$ to that of ${\bf D}_A^{-1}$, calculating ${\bf D}_C^{-1}{\bf E}_C{\bf D}_C^{-1}$ also takes $4K^2$ complex multiplications. As for the INS method, since ${\bf D}_I^{-1}{\bf E}_I{\bf D}_I^{-1}=({1}/{\omega_I^2}){\bf E}_I$, $K^2$ complex multiplications are needed for the second part.

For the TNS method, it has been reported in \cite{Prabhu_Hardware} that it needs $6K^2$ complex multiplications for the case of $L=1$. The explicit division number for ${\bf D}_T^{-1}$ was not provided. According to the classical Gauss-elimination method, about $K$ complex divisions are needed. In general, the inversion of the tri-diagonal matrix is not hardware-friendly, e.g., the modified Gauss-elimination-based algorithm used in \cite{Prabhu_Hardware} has the sequential nature which further reduces the computation efficiency of TNS \cite{Abbas_Low}. A summary of the above computation cost can be seen in Table I.
Among the considered four schemes, INS has the lowest complexity while TNS has the highest complexity. The complexity of ICNS is slightly lower than that of CNS. The complexity increase for ICNS is $3K^2$ complex multiplications compared with INS.

\begin{table}[H]\label{Table 1}
\vspace{-4mm}
\centering
\caption{Computational complexity of different precoding schemes}
\vspace{-5mm}
\begin{tabular}{|c|c|c|c|c|c|c|}
\hline
 & Multiplication & Division \\
\hline
INS & $K^2$ & 0\\
\hline
CNS &$4K^2$& $K$ \\
\hline
TNS &$6K^2$& $K$ \\
\hline
ICNS &$4K^2$  &$0$\\
\hline
\end{tabular}
\vspace{-4mm}
\end{table}

\subsection{Ordered ICNS Method}
Another improvement on ICNS can be obtained via exploiting the multi-user diversity gain, i.e., choosing the user with largest interference to others as User C. Therefore, the column with the largest 2-norm (excluding the diagonal elements in each column) is selected to make up the precondition matrix rather than the first column of $\bf G$. We name this ordered ICNS method. Mathematically, define $\tilde{\bf G}={\bf G}-{\rm diag}_0({\bf G})$, the precondition matrix ${\bf D}_B$ is given as
\begin{equation}\label{precondition_B}
{\bf D}_B=\omega{\bf I}_K+ \tilde{\bf G}_{j^\star},
\end{equation}
where $j^\star=\arg \max_{j}||\tilde{\bf g}_{j}||_F^2$ with $\tilde{\bf g}_{j}$ being the $j$th column of $\tilde{\bf G}$, $[\tilde{\bf G}_{j^\star}]_{i,{j^\star}}=[\tilde{\bf G}]_{i,{j^\star}}$ and $[\tilde{\bf G}_{j^\star}]_{i,j}=0, \forall j\neq{j^\star}, i$.

For the ordered ICNS, the norm calculation of all columns of $\tilde{\bf G}$ needs about $K^2$ complex multiplications. The \textit{max} operation has significantly lower complexity which can be omitted. Therefore, the computational complexity of the ordered ICNS is higher than that of ICNS by $K^2$ complex multiplications.

\section{Performance Analysis}\label{performance analysis}
In existing works, two criteria have been used to evaluate the performance of the low-complexity precoding designs \cite{Zhu_inversion}. The first one is the asymptotic convergence speed, i.e., {\small$\lambda_{max}(-{\bf D}^{-1}{\bf E})$}. The second one is the mean square error (MSE) between the noiseless received signals with the ideal ZF precoding and that with the approximate ZF procoding, i.e.,
\[{\rm E}\left\{{\left\|{\bf H}^H{\bf H}\left({\bf G}^{-1}-\sum\nolimits_{n=0}^{L}(-{\bf D}^{-1}{\bf E})^n{\bf D}^{-1}\right){\bf s}\right\|_F^2}\right\}.\]
These are both indirect metrics for the network performance. In this section, we work on the direct sum-rate performance. As there have been no sum-rate results for any of the aforementioned schemes, we conduct derivations for both the INS scheme as the most competitive benchmark for comparison, then for the proposed ICNS scheme. The method we use for performance analysis can be applied to other NS based low-complexity schemes.

\subsection{Sum-Rate Performance of INS}
Based on simulation results, we found that among all existing methods, INS is the most competitive one for comparison in terms of the tradeoff between performance and complexity. Therefore, we conduct its performance analysis for analytical comparison.

From \eqref{general_lowP} and \eqref{D_INS_inv}, the precoding matrix for the INS scheme can be written as
\begin{eqnarray}\label{W-INS}
{{\bf{W}}_I} = \frac{{{\beta_I}}}{M}{\bf{H}}\left( {\frac{2}{\omega }{{\bf{I}}_K} - \frac{1}{{{\omega ^2}}}{\bf{G}}} \right),
\end{eqnarray}
where $\beta_I$ is the power normalization parameter for ${{\bf{W}}_I}$. Notice that $\omega_I$ is replaced with $\omega$ for better presentation. Consequently, the equivalent channel matrix for the INS precoding can be represented as
\begin{eqnarray}\label{H_BETA}
  \hspace{-8pt}\tilde{\bf H}_I={\bf H}^H{\bf W}_{I} =\beta_I\left(\frac{2}{\omega}{\bf G}-\frac{1}{\omega^2}{\bf G}^2\right).
\end{eqnarray}
By drawing lessons from \cite[Lemma 1]{Zhang_Power}, we have the following analysis on
the ergodic sum-rate for large $M$ in massive MIMO systems,
\begin{eqnarray}\label{sum-rate_a_INS_def}
\small
\nonumber R_{sum}^{INS} \hspace{-3pt}&=&\hspace{-3pt}\sum_{k=1}^{K}{\rm E}\left\{\log_2\left(1+\frac{|[{\tilde{\bf H}}_I]_{kk}|^2}{\frac{1}{\rho_t}+\sum_{j\neq k}|[{\tilde{\bf H}}_I]_{kj}|^2}\right)\right\}\\
\hspace{-3pt}&\approx&\hspace{-3pt} \sum_{k=1}^{K}\log_2\left(1+\frac{{\rm E}\{|[{\tilde{\bf H}}_I]_{kk}|^2\}}{\frac{1}{\rho_t}+\sum_{j\neq k}{\rm E}\{|[{\tilde{\bf H}}_I]_{kj}|^2\}}\right).
\end{eqnarray}
A closed-form sum-rate approximation of the INS scheme is given in the following theorem.
\begin{theorem}\label{theorem 1}
For massive MIMO systems with the BS antenna number $M$, the channel correlation $c$, the user number $K$ and the operation SNR $\rho_t$, when $M\gg1$, the sum-rate of the INS precoding can be approximated as
\begin{equation}\label{sum-rate_INS}
R_{sum}^{INS}\approx K\log_2\left(1+\frac{C_1}{\frac{1}{\rho_t}\frac{K}{M}C_2+(K-1)C_3}\right),
\end{equation}
where
\begin{eqnarray}\label{eq39}
C_1\hspace{-0.1cm}=\hspace{-0.1cm}{{{\left(\hspace{-0.1cm} {2 - \frac{{\rm{1}}}{\omega }} \hspace{-0.1cm}\right)}^2}\hspace{-0.1cm}{\rm{ + }}\frac{4}{{c M}}{{\left(\hspace{-0.1cm} {1 - \frac{{\rm{1}}}{\omega }} \hspace{-0.1cm}\right)}^2} \hspace{-0.2cm}- \hspace{-0.1cm}\frac{2K}{{c M\omega}}\hspace{-0.1cm}\left(\hspace{-0.1cm} {2 - \frac{{\rm{1}}}{\omega }} \hspace{-0.1cm}\right)\hspace{-0.1cm} +\hspace{-0.1cm} \frac{K}{{{{c^2 M}^2\omega }}}\hspace{-0.1cm}\left(\hspace{-0.1cm} { - 4 + \frac{5}{\omega }} \hspace{-0.1cm}\right)\hspace{-0.1cm} +\hspace{-0.1cm} \frac{{{K^2}}}{{{{c^2 M}^2\omega ^2}}} \hspace{-0.1cm}+\hspace{-0.1cm} \frac{{{K^2}}}{{{{c^3 M}^3\omega ^2}}}},
\end{eqnarray}
\begin{equation}\label{eq41}
C_2={{{\left( {2 - \frac{1}{\omega }} \right)}^2} + \frac{K}{{c M\omega}}\left( { - 4 + \frac{3}{\omega }} \right) + \frac{{{K^2}}}{{{{c^2 M}^2\omega ^2}}}},
\end{equation}
and
\begin{equation}\label{eq40}
C_3={\frac{{\rm{4}}}{{c M}}{{\left( {1 - \frac{{\rm{1}}}{\omega }} \right)}^2}{\rm{ + }}\frac{K}{{{{c^2 M}^2\omega}}}\left( { - 4{\rm{ + }}\frac{5}{\omega }} \right){\rm{ + }}\frac{{{K^2}}}{{{{c^3 M}^3\omega ^2}}}}.
\end{equation}
\end{theorem}
\begin{IEEEproof}
See Appendix B. 
\end{IEEEproof}

Notice that one typical massive MIMO scenario is when $K$ increases with $M$ with a fixed ratio. The ${\mathcal O}(1/{M})$ terms in \eqref{eq39}-\eqref{eq40} are kept due to the multiplication coefficient $K$ in \eqref{sum-rate_INS}.

\subsection{Sum-Rate of the Proposed ICNS Scheme }\label{sr_icns}
From \eqref{precoding}, the equivalent channel matrix for the ICNS precoding is
\begin{eqnarray}
{\tilde{\bf H}_A}={\bf H}^H{\bf W}_{A}= \beta_A\left(2{\bf G}{\bf D}_A^{-1}-\left({\bf G}{\bf D}_A^{-1}\right)^2\right).
\end{eqnarray}
By following similar procedures in the sum-rate derivations for the INS precoding, but with a lot more involved details, a closed-form sum-rate approximation for the proposed ICNS scheme is given in the following theorem.
\begin{theorem}\label{theorem 2}
For massive MIMO systems with the BS antenna number $M$, the channel correlation $c$, the user number $K$ and the operation SNR $\rho_t$, when $M\ge K \gg1$, the sum-rate of the proposed ICNS precoding can be approximated as
\begin{eqnarray}\label{sum-rate-A}
R_{sum}^{A} \approx \underbrace{{\log _2}\left( {1 + \frac{C_4}{{\frac{1}{\rho_t}\frac{K}{M}C_5 + C_6}}} \right)}_{\text{Rate of User $1$}}+(K-1)\underbrace{{\log_2}\left(1+\frac{C_7}{{\frac{1}{\rho_t}\frac{K}{M}C_5 + C_8 }}\right)}_{\text{Rate of User $k$, $k=2,...,K$}}
\end{eqnarray}
where
\begin{eqnarray}\label{eq47}
\hspace{-5cm}C_4={\left( {2 - \frac{{\rm{1}}}{\omega } + \frac{3K}{{c M\omega}}\left( { - 1 + \frac{1}{\omega }} \right) + \frac{{{K^2}}}{{{{c^2 M}^2}\omega ^2}}\left( {1 - \frac{1}{\omega }} \right)} \right)^2},
\end{eqnarray}
\begin{eqnarray}\label{beta2}\nonumber
&&\hspace{-1.4cm}C_5\hspace{-0.1cm}=\hspace{-0.1cm}{\left( {2 - \frac{1}{\omega }} \right)^2} + \frac{K}{{c M\omega}}\left( { - 4{\rm{ + }}\frac{3}{\omega }} \right) + \frac{{{K^2}}}{{{{c^2 M}^2\omega ^2}}}+ \frac{2}{{c M\omega}}\left( { - 4 + \frac{{14}}{\omega } - \frac{{11}}{{{\omega ^2}}} + \frac{{\rm{2}}}{{{\omega ^3}}}} \right)\\
&&\hspace{-1.4cm}+\frac{K}{{{{c^2 M}^2\omega ^2}}}\left( {16 -\hspace{-0.1cm} \frac{{44}}{\omega } + \frac{{27}}{{{\omega ^2}}} - \frac{{\rm{4}}}{{{\omega ^3}}}} \right)\hspace{-0.1cm}+\hspace{-0.1cm} \frac{{{K^2}}}{{{{c^3 M}^3\omega ^3}}}\left( { \hspace{-0.1cm}- 4 + \frac{{13}}{\omega } - \frac{8}{{{\omega ^2}}} + \frac{{\rm{1}}}{{{\omega ^3}}}} \right) \hspace{-0.1cm}+\hspace{-0.1cm} \frac{{{K^3}}}{{{{c^4 M}^4\omega ^4}}}{\left( {1 \hspace{-0.1cm}-\hspace{-0.1cm} \frac{1}{\omega }} \right)^2},
\end{eqnarray}
\begin{eqnarray}\label{eq48}
\hspace{-2.2cm}{C_6} = \frac{4K}{{c M}}{\left( {{\rm{1}} - \frac{{\rm{1}}}{\omega }} \right)^2} - \frac{{{K^2}}}{{{{c^2 M}^2\omega}}}\left( {4{{\left( {{\rm{1}} - \frac{1}{\omega }} \right)}^2} - \frac{1}{\omega }} \right) + \frac{{{K^3}}}{{{{c^3 M}^3\omega ^2}}}{\left( {1 - \frac{1}{\omega }} \right)^2},
\end{eqnarray}
\begin{eqnarray}\label{eq49}\nonumber
\hspace{-4.4cm}C_7\hspace{-0.2cm}&=&\hspace{-0.2cm}{\left(\hspace{-0.1cm} {2 - \frac{{\rm{1}}}{\omega }} \hspace{-0.1cm}\right)^2} \hspace{-0.2cm}- \hspace{-0.1cm} \frac{{{\rm{2}}K}}{cM\omega }\hspace{-0.1cm}\left(\hspace{-0.1cm} {2 - \frac{{\rm{1}}}{\omega }} \hspace{-0.1cm}\right) \hspace{-0.1cm}+\hspace{-0.1cm} \frac{{{K^2}}}{{{c^2M^2\omega ^2}}}{\rm{ + }}\frac{2}{{c M}}\hspace{-0.1cm}\left(\hspace{-0.1cm} {2 - \frac{{\rm{4}}}{\omega }{\rm{ + }}\frac{4}{{{\omega ^2}}} - \frac{{\rm{1}}}{{{\omega ^3}}}}\hspace{-0.1cm} \right)\\
 \hspace{-4.4cm}&+&\hspace{-0.2cm} \frac{K}{c^2 M^2\omega }\left( { - 4 + \frac{9}{\omega } - \frac{4}{{{\omega ^2}}}} \right) + \frac{K^2}{{{c^3 M^3\omega ^2}}}\left( {1 - \frac{2}{\omega }} \right),
\end{eqnarray}
and
\begin{eqnarray}\label{eq50}\nonumber
\hspace{-1.2cm}C_8\hspace{-0.2cm}&=&\hspace{-0.2cm}\frac{4K}{{c M}}{\left( {1 - \frac{{\rm{1}}}{\omega }} \right)^2} + \frac{{{K^2}}}{{{{c^2 M}^2\omega}}}\left( { - 4 + \frac{5}{\omega }} \right) + \frac{{{K^3}}}{{{{c^3 M}^3\omega ^2}}} + \frac{1}{{c M}}\left( {\frac{4}{{{\omega ^2}}}{{\left( { - 2 + \frac{{\rm{1}}}{\omega }} \right)}^2} - 4} \right)\\
\nonumber\hspace{-1.2cm}&+&\hspace{-0.2cm}\frac{K}{{{{c^2 M}^2\omega}}}\left( { - 4 + \frac{{{\rm{51}}}}{\omega } - \frac{{84}}{{{\omega ^2}}} + \frac{{{\rm{38}}}}{{{\omega ^3}}} - \frac{{\rm{4}}}{{{\omega ^4}}}} \right) + \frac{{{K^2}}}{{{{c^3 M}^3\omega ^2}}}\left( {15 - \frac{{66}}{\omega } + \frac{{{\rm{65}}}}{{{\omega ^2}}} - \frac{{{\rm{20}}}}{{{\omega ^3}}} + \frac{{\rm{1}}}{{{\omega ^4}}}} \right)\\
\hspace{-1.2cm}&+&\hspace{-0.2cm}\frac{{{K^3}}}{{{{c^4 M}^4\omega ^3}}}\left( { - 8 + \frac{{21}}{\omega } - \frac{{{\rm{16}}}}{{{\omega ^2}}} + \frac{{\rm{3}}}{{{\omega ^3}}}} \right) + \frac{{{K^4}}}{{{{c^5 M}^5\omega ^4}}}{\left( {1 - \frac{{\rm{1}}}{\omega }} \right)^2}.
\end{eqnarray}
\end{theorem}
\begin{IEEEproof}
See Appendix C. 
\end{IEEEproof}

Note that for the effective SINR of User 1, the lower order ${\mathcal O}(1/{ M})$ terms in the signal power in \eqref{eq47} and interference power in \eqref{eq48} are omitted. However, ${\mathcal O}(1/{ M})$ terms are kept in those of the effective SINR of Users 2 to $K$ due to the multiplication coefficient $K-1$ in \eqref{sum-rate-A}. Also, for ICNS, while Users 2 to $K$ have the same effective SINR, the effective SINR of User $1$ is different due to the consideration of User $1$'s interference to others in the precondition matrix design. This is different to INS, where the users are treated homogeneously.

\subsubsection{The Effect of $\omega$ on the Sum-Rate}
\begin{figure*}[t]
  \normalsize
  \centering
  \hspace{-0.9cm}
  \begin{minipage}[t]{0.48\textwidth}
    \centering
    \includegraphics[scale=0.6]{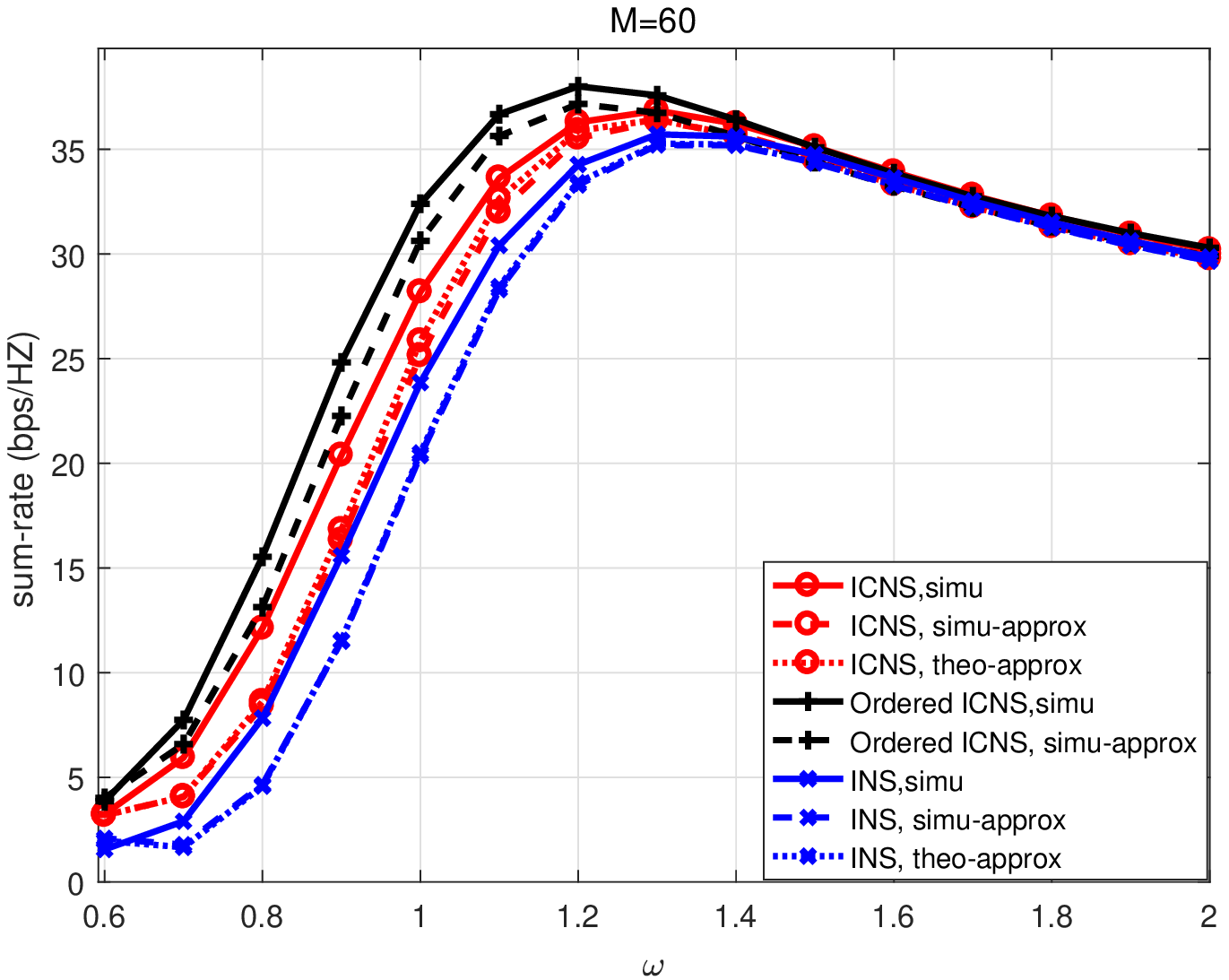}\vspace{-1mm}
    \vspace{-1mm}
  \end{minipage}
  \hspace{0.3cm}
  \begin{minipage}[t]{0.48\textwidth}
    \centering
    \includegraphics[scale=0.6]{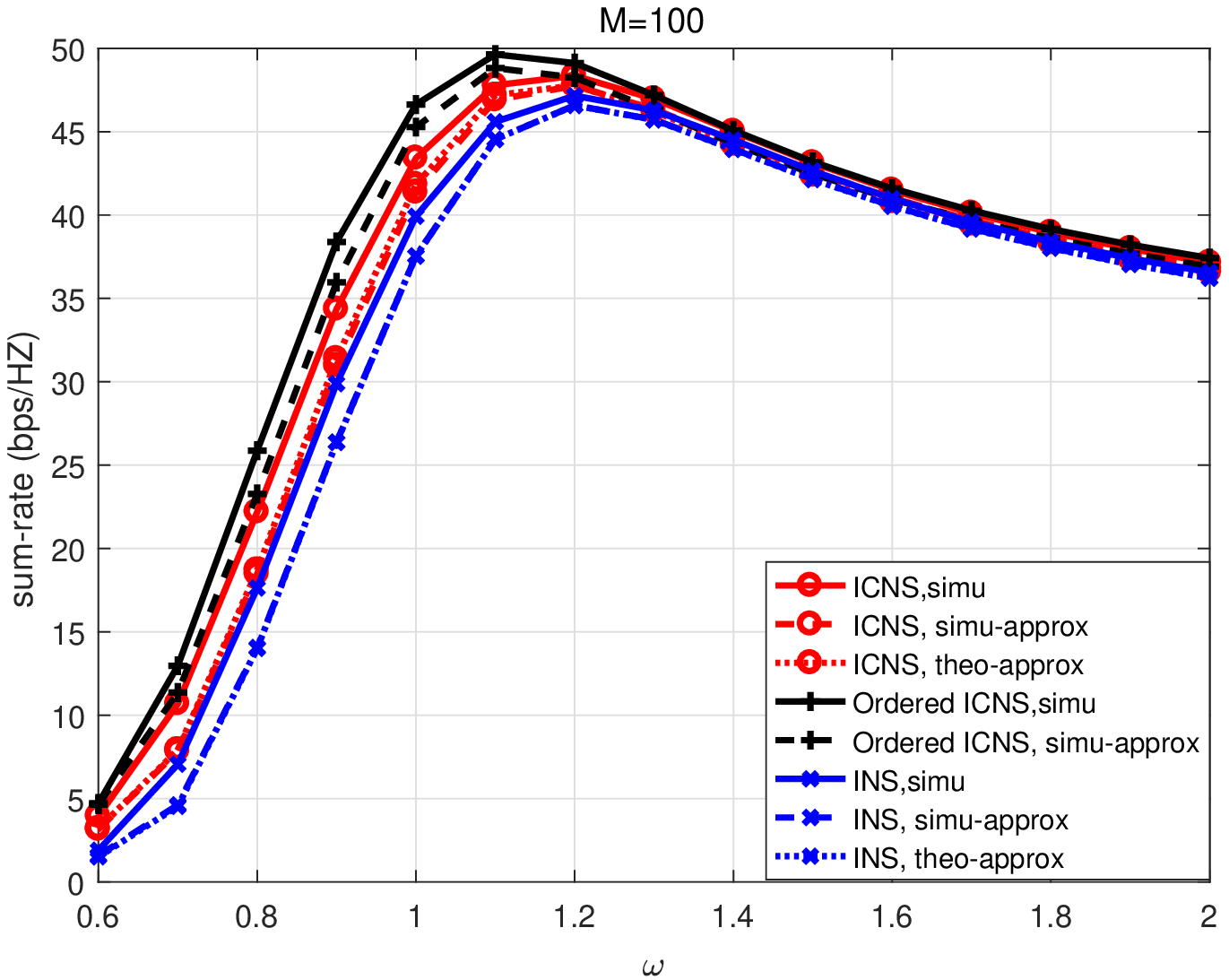}\vspace{-1mm} 
  \end{minipage}
  \vspace{-6mm}
  \caption{The effect of $\omega$ on the sum-rate of INS, ICNS and the ordered ICNS. $K=10$, $c=0.5$, $\rho_t=10$. Left: $M=60$. Right: $M=100$. ``simu" denotes the simulated sum-rate; ``simu-approx" denotes the simulated sum-rate approximation defined in \eqref{sum-rate_a_INS_def}; ``theo-approx" denotes the theoretical sum-rate approximation given in Theorem \ref{theorem 1} and \ref{theorem 2}.}\label{omega_M60_2}
  \vspace{-9mm}
\end{figure*}

With the above derived closed-form sum-rate approximations, we can study the effect of $\omega$ on the sum-rate performance and solve the optimal $\omega$ for the INS scheme and the proposed ICNS scheme, respectively, via one-dimensional grid search for given channel correlation level $c$, the BS antenna number $M$, the user number $K$ and the operation SNR $\rho_t$. In Fig.~\ref{omega_M60_2}, the sum-rates of the INS, the proposed ICNS and ordered ICNS schemes are shown where $c=0.5$, $K=10$, $\rho_t=10$ and $M=60$ or $100$.
For $M=60$, the optimal $\omega$ values of INS and ICNS are both $1.3$ while that of the ordered ICNS is $1.2$. For $M=100$, the optimal $\omega$ values of INS and ICNS are both $1.2$ while that of the ordered ICNS is $1.1$. The heuristic values, $\omega^{\star}$ in Lemma \ref{Lemma 1}, for $M=60$ and $M=100$ are $1.33$ and $1.2$, respectively.
First, the heuristic value is close to the optimal one, especially for INS and ICNS. Meanwhile,
this difference between $\omega^{\star}$ and the optimal $\omega$ only results in small performance degradation. The plots also show that for $\omega\ge\omega^{\star}$ the gap between the sum-rates of proposed ICNS/ordered ICNS and that of INS is relatively small, while for $\omega<\omega^{\star}$ this gap is larger. This is because that for large $\omega$, the effect of extra non-zero and non-diagonal elements in the precondition matrices of ICNS/ordered ICNS become negligible compared with the diagonal elements.

\subsection{Performance Comparison of MRT, ZF, INS and Proposed ICNS}\label{performance_comparison}
In the section, we compare the sum-rates of MRT, ZF, INS and the proposed ICNS. Note that the relaxation parameters in INS and ICNS are the same as given in \eqref{heuristic_omega}. Three typical cases are considered: 1) asymptotically large BS antenna number $M$ and user number $K$ with a fixed loading factor $r$; 2) finite $K$ (e.g., $K=10$ as typical value) with large but finite $M$; 3) finite $K$ with asymptotically large $M$. Note that Case 3 is a special case of Case 2.

The sum-rates of the ideal ZF and MRT precodings are given first.
With the ideal ZF precoding in \eqref{perfect ZF} and the power normalization in \eqref{eq3_1},
the sum-rate of the ideal ZF is
\begin{equation}\label{sum-rate-ZF}
R_{sum}^{ZF}=K\log_2\left(1+\widetilde{{\rm SINR}}_{ZF}\right)=K\log_2\left(1+\rho_t\left(\frac{M}{K}-\frac{1}{c}\right)\right),
\end{equation}
where the effective SINR of User $k$ follows from
\begin{eqnarray}\label{sinr_ZF}
\widetilde{{\rm SINR}}_{ZF}=\rho_t\beta_{ZF}^2= \frac{\rho_t}{{\rm E}\{{\rm tr}\{({\bf H}^H{\bf H})^{-1}\}\}}=
\frac{\rho_t}{c{\rm E}\{{\rm tr}\{({\bf \tilde Z}^H{\bf \tilde Z})^{-1}\}\}}=\rho_t\frac{c{M}-K}{cK},
\end{eqnarray}
where $\tilde{\bf Z}={\bf A}^H{\bf Z}$ is a $cM\times K$ matrix with each column following $\mathcal{CN}(0,{\bf I}_{cM})$ independently and the last equality follows from the property of the central complex Wishart matrix \cite{Tulino}.

By drawing lessons from \cite{Cheng_Performance}, a tight sum-rate lower bound of MRT can be expressed as
\begin{equation}\label{sum-rate-MRT}
R_{sum}^{MRT}\geq K\log_2\left(1+\widetilde{{\rm SINR}}_{MRT}\right)=K\log_2\left(1+{M}{\big/}\left({\frac{K-1}{c}+\frac{K}{\rho_t}}\right)\right).
\end{equation}

\subsubsection{Asymptotically Large ${M}$ and $K$ with a Fixed Ratio $r$}
Since the maximum multiplexing gain in the channel with correlation level $c$ is $cM$, the practical range of $r$ is $(0,c]$.
In the sum-rate expression for INS in \eqref{sum-rate_INS}, $C_1$ represents the normalized signal power, $\frac{1}{\rho_t}\frac{K}{M}C_2$ represents the normalized noise power, and $(K-1)C_3$ represents the normalized interference power.
Similarly, in the sum-rate expression for ICNS in \eqref{sum-rate-A}, $C_4$ and $C_7$ represent the normalized signal power and $C_6$ and $C_8$ represent the normalized interference power for User $1$ and User $2,\cdots,K$, respectively. $\frac{1}{\rho_t} \frac{K}{M}C_5$ represents the normalized noise power for both User $1$ and User $2,\cdots,K$.

First, the normalized noise power in the SINR of ICNS and that of INS are compared.
\begin{eqnarray}
\nonumber\hspace{-0.5cm}\frac{1}{{{\rho _t}}}\frac{K}{M}{C_5} \hspace{-0.1cm}-\hspace{-0.1cm} \frac{1}{{{\rho _t}}}\frac{K}{M}{C_2} \hspace{-0.3cm}&=& \hspace{-0.3cm} \frac{1}{{{\rho _t}}}r{\bigg [}{\frac{2}{{c M\omega}}\left( { - 4 + \frac{{14}}{\omega } - \frac{{11}}{{{\omega ^2}}} + \frac{{\rm{2}}}{{{\omega ^3}}}} \right) \hspace{-0.1cm}+\hspace{-0.1cm} \frac{r}{{{{c^2 M}\omega ^2}}}\left( {16 - \frac{{44}}{\omega } + \frac{{27}}{{{\omega ^2}}} - \frac{{\rm{4}}}{{{\omega ^3}}}} \right)}\\
&+&\hspace{-0.4cm} \frac{{{r^2}}}{{{{c^3 M}\omega ^3}}}\hspace{-0.1cm}\left( { - 4 + \frac{{13}}{\omega } - \frac{8}{{{\omega ^2}}} + \frac{{\rm{1}}}{{{\omega ^3}}}} \right) \hspace{-0.1cm}+\hspace{-0.1cm} \frac{{{r^3}}}{{{{c^4 M}\omega ^4}}}{{\left( {1 - \frac{1}{\omega }} \right)}^2}{\bigg ]}\hspace{-0.1cm}=\hspace{-0.1cm}\mathcal{O}\left(\frac{1}{\rho_tM}\right)\hspace{-0.1cm}.
\end{eqnarray}

\begin{figure*}[t]
  \normalsize
  \centering
  \hspace{-0.9cm}
  \begin{minipage}[t]{0.48\textwidth}
    \centering
    \includegraphics[scale=0.6]{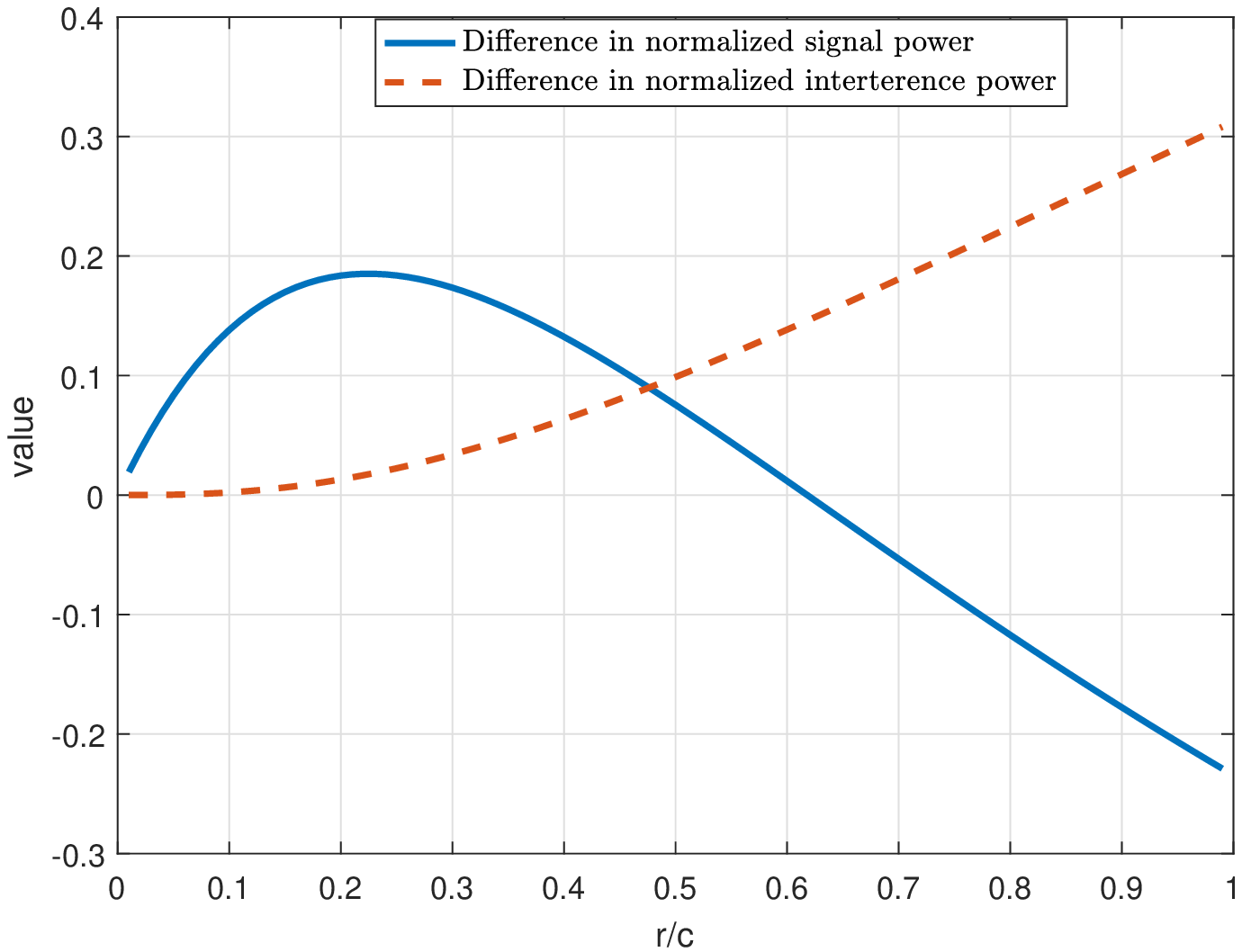}\vspace{-1mm}
    \vspace{-1mm}
  \end{minipage}
  \hspace{0.3cm}
  \begin{minipage}[t]{0.48\textwidth}
    \centering
    \includegraphics[scale=0.6]{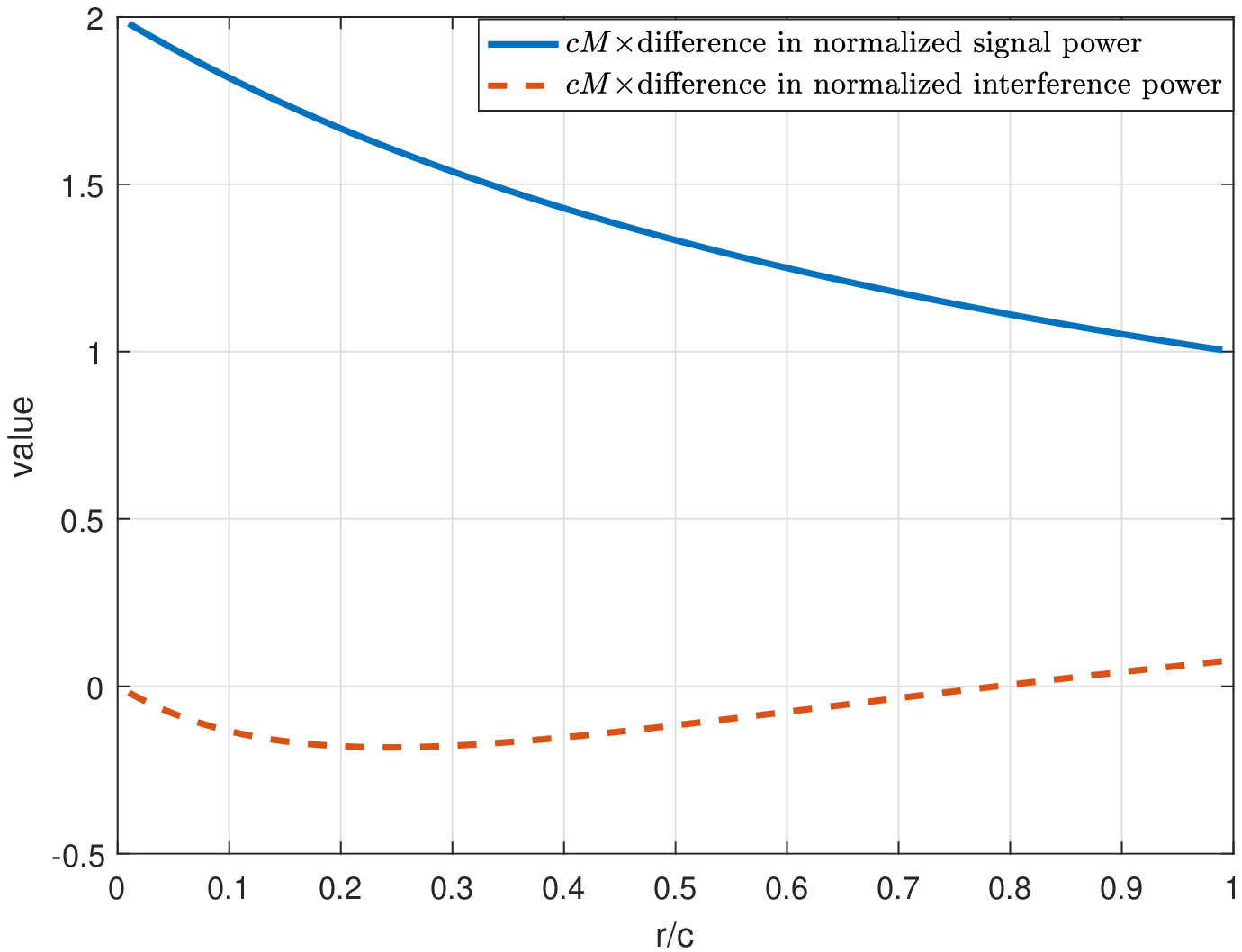}\vspace{-1mm}
  \end{minipage}
  \vspace{-7mm}
  \caption{The gap between the normalized signal/interference power of ICNS and INS for User 1 (left) and $k=2,...,K$ (right).}\label{delta_SINR1_int}
  \vspace{-10mm}
\end{figure*}

Comparisons of the normalized signal power and interference power are then conducted for User $1$ and User $k=2,...,K$, separately, due to their different forms in the ICNS scheme. For User $1$, the gap between the normalized signal power of ICNS and INS is
\begin{eqnarray}\label{user1_sig_gap}
{C_4} \hspace{-0.1cm}-\hspace{-0.1cm} {C_1} \hspace{-0.1cm}=\hspace{-0.1cm} \frac{r}{c}\hspace{-0.1cm}\left( { - \frac{8}{\omega } + \frac{{16}}{{{\omega ^2}}} - \frac{6}{{{\omega ^3}}}} \right) \hspace{-0.1cm}+\hspace{-0.1cm} \frac{{r}^2}{c^2}\hspace{-0.1cm}\left( {\frac{{12}}{{{\omega ^2}}} - \frac{{24}}{{{\omega ^3}}} \hspace{-0.1cm}+\hspace{-0.1cm} \frac{{11}}{{{\omega ^4}}}} \right) \hspace{-0.1cm}+ \hspace{-0.1cm}\left(\hspace{-0.1cm} { - \frac{{6{{r}^3}}}{{{c^3\omega ^3}}} + \frac{{{{ r}^4}}}{{{c^4\omega ^4}}}} \right)\hspace{-0.2cm}{\left(\hspace{-0.1cm} {1 \hspace{-0.1cm}-\hspace{-0.1cm} \frac{1}{\omega }} \right)^2}\hspace{-0.1cm}+\hspace{-0.1cm}\mathcal{O}\hspace{-0.1cm}\left(\frac{1}{M}\right),
\end{eqnarray}
and the gap between the normalized interference power is
\begin{eqnarray}\label{user1_int_gap}
{C_6} - \left( {K - 1} \right){C_3} = \frac{{4{{ r}^2}}}{{{c^2\omega ^2}}}\left( {1 - \frac{1}{\omega }} \right) + \frac{{{{ r}^3}}}{{{c^3\omega ^3}}}\left( {\frac{1}{\omega } - 2} \right)+\mathcal{O}\left(\frac{1}{M}\right).
\end{eqnarray}
Recall that $\omega=\omega^{\star}=1+{r}/c$. The values of the two gaps in \eqref{user1_sig_gap} and \eqref{user1_int_gap} with respect to ${r}/c$ are shown in the left sub-figure of Fig. \ref{delta_SINR1_int}, where the effective loading factor ${r}/c$ is used for better clarification.
It can been seen that for User 1, the normalized signal power of ICNS is larger than that of INS when ${r}\le0.61c$; while as $r$ further increases the gap decreases to a negative value. On the other hand, the normalized interference power of ICNS is always larger than that of INS and the gap increases as ${ r}$ increases.

For $\rho_t\gg1$, which is favorable for ZF-like precodings, the gap between the normalized noise power of ICNS and INS can be ignored.
Therefore, for ${ r}>0.61c$, the effective SINR of User 1 with ICNS is smaller than that with INS. Further for $r\in[0.22c,0.61c]$, as ${ r}$ decreases, the effective SINR of User 1 with ICNS approaches or even surpass that with INS due to its larger signal power increment and smaller interference power increment compared with INS.

For User $k=2,...,K$, the gap between the normalized signal power of ICNS and INS is
\begin{eqnarray}
{C_7} - {C_1} = \frac{1}{{c M}}\frac{{\rm{2}}}{{{\omega ^2}}}\left( {2 - \frac{{\rm{1}}}{\omega }{\rm{ + }}\frac{2 r}{c}\left( {1 - \frac{1}{\omega }} \right) - \frac{{{{ r}^2}}}{c^2\omega }} \right),
\end{eqnarray}
and the gap between the normalized interference power is
\begin{eqnarray}
\nonumber&&\hspace{-0.95cm}{C_8}  \hspace{-0.1cm}- \hspace{-0.1cm} \left( {K - 1} \right){C_3} \hspace{-0.1cm}= \hspace{-0.15cm} \frac{1}{{c M}}{\bigg [}{\frac{4}{\omega }\left( 1-\frac{1}{\omega} \right)^2\hspace{-0.2cm}\left( { - 2 + \frac{{\rm{1}}}{\omega }} \right) \hspace{-0.1cm}+  \hspace{-0.1cm} \frac{{\rm{1}}}{\omega }\left( { - 8 + \frac{{{\rm{56}}}}{\omega } - \frac{{84}}{{{\omega ^2}}} + \frac{{{\rm{38}}}}{{{\omega ^3}}} - \frac{{\rm{4}}}{{{\omega ^4}}}} \right)\frac{r}{c}}\\
&&\hspace{-1.4cm}+{\frac{1}{{{\omega ^2}}}\left( {16 - \frac{{66}}{\omega } + \frac{{{\rm{65}}}}{{{\omega ^2}}} - \frac{{{\rm{20}}}}{{{\omega ^3}}}  \hspace{-0.1cm}+ \hspace{-0.1cm} \frac{{\rm{1}}}{{{\omega ^4}}}} \right){\frac{{ r}^2}{c^2}} + \frac{1}{{{\omega ^3}}}\left( { - 8 + \frac{{21}}{\omega } - \frac{{{\rm{16}}}}{{{\omega ^2}}} + \frac{{\rm{3}}}{{{\omega ^3}}}} \right){\frac{{ r}^3}{c^3}} \hspace{-0.1cm}+ \hspace{-0.1cm} \frac{{\rm{1}}}{{{\omega ^4}}}{{\left( {1 - \frac{{\rm{1}}}{\omega }} \right)}^2}{\frac{{ r}^4}{c^4}}}{\bigg ]}.
\end{eqnarray}
The normalized values of the two gaps (via multiplying by $c{M}$ to focus on the effect of ${r}/c$) are shown in the right sub-figure of Fig. \ref{delta_SINR1_int}. It can be seen that for User $k=2,...,K$, ICNS results in larger signal power for the whole ${ r}$ range and smaller interference power for ${ r}\le 0.8c$. For ${ r}>0.8c$, ICNS brings slightly higher interference power. Recall that the gap between the normalized noise power of ICNS and INS can be ignored for $\rho_t\gg1$. Therefore, for ${ r}\le 0.8c$, ICNS results in larger effective SINR for User $k=2,...,K$.

\begin{remark}
Based on the above discussions, for the case of asymptotically large ${M}$ and $K$ with a fixed non-zero $r$ and $\rho_t\gg 1$, ICNS outperforms INS in sum-rate for ${ r}\in[0.22c,0.8c]$ due to the SINR increase for $K-1$ users. Moreover, the advantage is larger for small $r$. This is because that as ${r}$ decreases, i.e., smaller $K$ for any given $M$, the ratio of the number of interference terms that are considered in ICNS, i.e., $K-1$, to the whole number of interference terms, i.e., $K^2-K$, becomes larger. For ${ r}>0.8c$, the sum-rate gap between ICNS and INS decreases to some extent.
For ${ r}<0.22c$, ICNS may still have higher sum-rate than INS, while for ${ r}\rightarrow 0$, since the Gram matrix approaches the identity matrix, ICNS and INS both approaches ZF precoding and thus have the same performance.
\end{remark}

Next, we derive the favorable $r$ range of the INS and ICNS, i.e., the range of $r$ that makes their sum-rates approach or even surpass that of the ideal ZF and no worse than that of MRT simultaneously. The second condition follows from that for certain large $r$, even MRT can outperform ZF in terms of sum-rate due to the large cost of degrees of freedom for interference cancellation in ZF.
Since for the proposed ICNS, the SINR of User 1 is different from those of Users $2$ to $K$, we study the above problem with the help of the analytical results on INS and deduce the conclusion for ICNS based on their relationship. First we give the following corollary.

\begin{corollary}\label{corollary1}
For massive MIMO systems with channel correlation level $c$, SNR $\rho_t$ and asymptotically larger $M$ and $K$ with fixed ratio $r$, the sum-rate of INS is larger than that of MRT when $\rho_t>rc/(r+c)$ and approximates that of ideal ZF when $r$ equals to
\begin{equation}\label{pos_root}
{r}^*=\frac{{\sqrt {9{c^2} + 4c{\rho _t} + 4\rho _t^2}  - 3c}}{{2\left( {c + {\rho _t}} \right)}}c.
\end{equation}
\end{corollary}
\begin{IEEEproof}
See Appendix D. 
\end{IEEEproof}

\begin{figure}[t]
\centering
\includegraphics[scale=0.6]{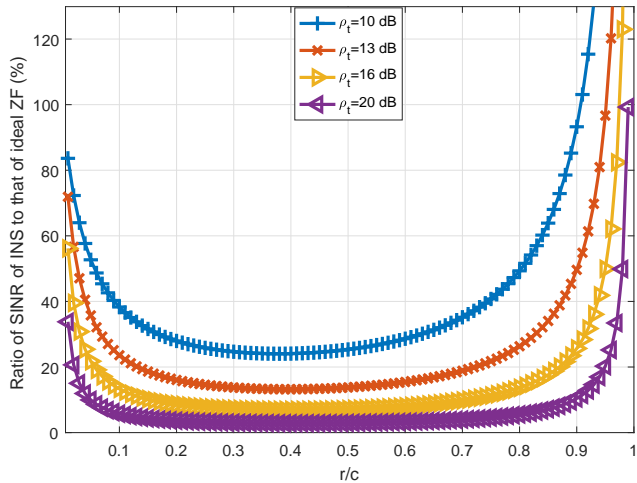}
\vspace{-6mm}
\caption{The effect of $r$ on the ratio of the SINR of INS to that of the ideal ZF. $c=0.5$}\label{effect_r_INS}
\vspace{-10mm}
\end{figure}

Since ${{ r}c}/({ r}+c)<1$, it can be known from Corollary \ref{corollary1} that INS has higher sum-rate than MRT for $\rho_t > 1$ (i.e., more than $0$ dB). Notice that ${r}^*<c$. Moreover, with the help of the following plots in Fig. \ref{effect_r_INS}, we know that the favorable $r$ range is $[{r}^*,c]$ if $\rho_t>1$.
Specifically, Fig. \ref{effect_r_INS} demonstrates the relationship between the ratio of the effective SINR of INS to that of the ideal ZF $Pr_{\rm INS}$ and $r$ where $c=0.5$, $\rho_t=10,13,16,20$ dB and the corresponding values of ${{r}^*}$ are $0.9071c$, $0.9517c$, $0.9753c$ and $0.9901c$, respectively.
It can be seen that 1) the closed-form expression for ${{r}^*}$ in \eqref{pos_root} is accurate; 2) INS has no smaller sum-rate than the ideal ZF for $r \in [{r}^*,c]$. Moreover, ${{r}^*}$ increases as $\rho_t$ increases.
Since the proposed ICNS has higher sum-rate than INS for $r\in[0.22c,0.8c]$ and $\rho_t\gg 1$ as discussed above, an conservative estimation of the favorable range $r$ of ICNS is about $[\max({r}^*,0.22c),0.8c]$ if ${r}^*\le0.8c$.

\subsubsection{Finite $K$ with Large but Finite $M$}\label{case2}
Now we consider large but finite ${M}$ and finite $K$ (e.g., $K=10$) which is the most general and practical case.
For the INS sum-rate result in Theorem \ref{theorem 1}, the terms with $M^2$ or higher order term in their denominators,
e.g., $K^2/{M}^2$, can be omitted in the effective SINR components, since they are lower order terms with respect to $M$ compared with the remaining terms with $1/M$ or $K/M$. The terms with $1/M$ are kept due to their non-negligible effect on the comparison for Users $k=2,...,K$.
Thus, from \eqref{eq39}-\eqref{eq40}, the SINR components for INS can be first approximated as
\begin{eqnarray}
\label{eq99}{C_1} \hspace{-0.2cm}&\approx&\hspace{-0.2cm} {\left( {2 - \frac{{\rm{1}}}{\omega }} \right)^2}{\rm{ + }}\frac{4}{{c M}}{\left( {1 - \frac{1}{\omega }} \right)^2} - \frac{{\rm{2}}}{\omega }\frac{K}{{c M}}\left( {2 - \frac{{\rm{1}}}{\omega }} \right),\\
  \label{eq98}\frac{1}{\rho_t}\frac{K}{M}C_2 \hspace{-0.2cm}&\approx&\hspace{-0.2cm} \frac{K}{{M{\rho _t}}}{\left( {2 - \frac{1}{\omega }} \right)^2},\\
\label{c_3_aa}\left( {K - 1} \right){C_3} \hspace{-0.2cm}&\approx&\hspace{-0.2cm} \frac{{\left( {K - 1} \right)}}{{c M}}{\rm{4}}{\left( {1 - \frac{{\rm{1}}}{\omega }} \right)^2}.
\end{eqnarray}
Similarly for ICNS whose sum-rate result is given in Theorem \ref{theorem 2}, the terms with $M^2$ or higher order term in their denominators are omitted in the effective SINR components for each user. For User 1, the terms with $1/M$ are omitted as well.
From \eqref{eq47}-\eqref{eq50}, we have
\begin{eqnarray}
\label{c_4_a}{C_4} \hspace{-0.2cm}&\approx&\hspace{-0.2cm} {\left( {2 - \frac{{\rm{1}}}{\omega }} \right)^2} - \frac{6}{\omega }\left( {2 - \frac{{\rm{1}}}{\omega }} \right)\left( {1 - \frac{1}{\omega }} \right)\frac{K}{{c M}},\\
 \frac{1}{{{\rho _t}}}\frac{K}{M}{C_5} \hspace{-0.2cm}&\approx&\hspace{-0.2cm} \frac{K}{{M{\rho _t}}}{\left( {2 - \frac{1}{\omega }} \right)^2}\approx\frac{1}{{{\rho _t}}}\frac{K}{M}{C_2},\\
\label{c_6_a}{C_6} \hspace{-0.2cm}&\approx&\hspace{-0.2cm}\frac{K}{{c M}}{\rm{4}}{\left( {1 - \frac{{\rm{1}}}{\omega }} \right)^2},\\
{C_7} \hspace{-0.2cm}&\approx&\hspace{-0.2cm} {\left( {2 - \frac{{\rm{1}}}{\omega }} \right)^2}{\rm{ + }}\frac{4}{{c M}}{\left( {1 - \frac{1}{\omega }} \right)^2} - \frac{{\rm{2}}}{\omega }\frac{K}{{c M}}\left( {2 - \frac{{\rm{1}}}{\omega }} \right) + \frac{1}{{c M}}\frac{{\rm{2}}}{{{\omega ^2}}}\left( {2 - \frac{{\rm{1}}}{\omega }} \right),\\
\label{eq60}{C_8} \hspace{-0.2cm}&\approx&\hspace{-0.2cm} {\rm{4}}{\left( {1 - \frac{{\rm{1}}}{\omega }} \right)^2}\frac{K}{{c M}} + \left( {\frac{{\rm{4}}}{{{\omega ^2}}}{{\left( {\frac{{\rm{1}}}{\omega } - 2} \right)}^2} - 4} \right)\frac{1}{{c M}}.
\end{eqnarray}

To compare the effective SINR of User $1$ with ICNS and that with INS, we further neglect the terms with $1/M$ in the approximations of $C_1$ and $\left( {K - 1} \right){C_3}$ in \eqref{eq99} and \eqref{c_3_aa}, respectively. Thus, the gap of the normalized signal power and that of normalized interference power for ICNS and INS are
\begin{eqnarray}
  &&{C_4} - {C_1} \approx  - \frac{2}{{{\omega ^3}}}\frac{K}{{c M}}\left( {4{{\left( {\omega  - 1} \right)}^2} - 1} \right), \quad {C_6} - \left( {K - 1} \right){C_3} \approx 0,
\end{eqnarray}
respectively. Recall that $\omega=1+K/(cM)$ and $K/(cM) \ll 1$ for the considered case which means $w \approx 1$.
Since $-\frac{2}{{{\omega ^3}}}\frac{K}{{c M}}\left( {4{{\left( {\omega  - 1} \right)}^2} - 1} \right) = -\frac{2}{{{\omega ^3}}}\left( {\omega  - 1} \right)\left( {4{{\left( {\omega  - 1} \right)}^2} - 1} \right)>0$ for $\omega\in (1,1.5)$, it can be concluded that ICNS results in larger signal power than INS for User 1, and consequently larger effective SINR due to the same interference power.
Meanwhile, in the interval $\omega\in (1,1.5)$, ${C_4} - {C_1}$ first increases and then decreases as $\omega$ decreases (via the increase of $M$) where the maximum point is reached at $\omega=1.21$ (i.e., ${M}=K/(0.21c)$). Thus, as $M$ grows, the gap between the effective SINR of User $1$ with ICNS and that with INS first increases for relatively small $M$ and then decreases as $M$ further grows.

\begin{figure}[t]
\centering
\includegraphics[scale=0.6]{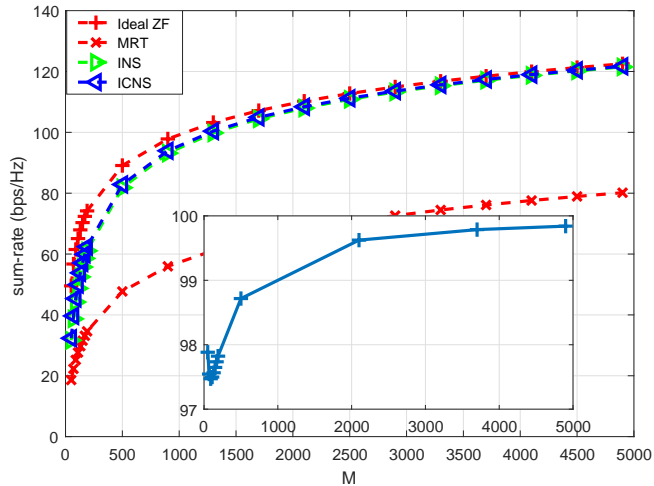}
\vspace{-7mm}
\caption{Comparison among ICNS, INS, MRT and the ideal ZF in terms of sum-rate for finite $K$ and growing large $M$. $c=0.5$, $\rho_t=10$ and $K=10$. The ratio of the sum-rate of INS to that of ICNS is shown in the sub-figure. }\label{fixed_K_finite or infiniely M}
\vspace{-10mm}
\end{figure}

To compare the effective SINR of User $k=2,...,K$ with ICNS and that with INS, from \eqref{eq99}-\eqref{eq60} we have ${C_7} - {C_1} \approx \frac{1}{{c M}}\frac{{\rm{2}}}{{{\omega ^2}}}\left( {2 - \frac{{\rm{1}}}{\omega }} \right) = \frac{{\omega  - 1}}{K}\frac{4}{{{\omega ^3}}}\left( {\omega  - \frac{1}{2}} \right) > 0$ for $\omega=1+K/(c{M})>1$ and it decreases to zero as $ M$ grows. Meanwhile,
${C_8} - \left( {K - 1} \right){C_3} \approx \frac{{\rm{4}}}{\omega }\left( {\frac{{\rm{1}}}{\omega } - 2} \right){\left( {1 - \frac{{\rm{1}}}{\omega }} \right)^2}\frac{{\omega  - 1}}{K}$ is negative and increases to zero as $M$ grows. Therefore, the effective SINR of User $k=2,...,K$ with ICNS is larger than that with INS and the gap decreases as $M$ grows large.
An example for the comparison between the sum-rate of ICNS and that of INS with finite $K$ and large but finite $M$ is given in Fig. \ref{fixed_K_finite or infiniely M} where $c=0.5$, $\rho_t=10$ and $K=10$. From the ratio of the sum-rate of INS to that of ICNS, it can be seen that the sum-rate of ICNS is better than that of INS and as $M$ grows, the gap first increases for relatively small $M$ and then decreases to zero.
The initial increasing trend is in accordance with the trend of SINR gap for User $1$, while the decreasing trend is intuitive, i.e., as $M$ grows large, the Gram matrix $\bf G$ approaches the identity matrix well and difference between ICNS and INS becomes negligible.

\subsubsection{Finite $K$ with Asymptotically Large $M$}
This is actually the asymptotic case of the above where $M$ can further grow infinitely. Correspondingly, $\omega=1+K/(c{M})\rightarrow1$. Based on the analysis for the above case, we know that with any given $K$, the sum-rates of ICNS and INS become the same as ${M}$ grows very large. Furthermore, they both grow to infinity as ${M}$ grows to infinity. This can be easily seen via further neglecting the terms with $K/{M}$ and $1/{M}$ in all SINR components of INS and ICNS in \eqref{eq99}-\eqref{eq60}.
In existing work, finer observations on the behavior of sum-rates of ideal ZF and MRT are based on the following approximations of \eqref{sum-rate-ZF} and \eqref{sum-rate-MRT}:
\begin{eqnarray}
R_{sum}^{ZF} \hspace{-0.2cm}&\approx&\hspace{-0.2cm} K\log_2\left(1+\frac{M\rho_t}{K}\right), \quad R_{sum}^{MRT} \approx K\log_2\left(1+\frac{M\rho_t}{\frac{(K-1)\rho_t}{c}+K}\right).
\end{eqnarray}
Using a similar approximation, i.e., $\omega\approx1$ and the terms with $K/{M}$ and $1/{M}$ and non-zero coefficient in \eqref{c_4_a}-\eqref{eq60} are kept intact, we have
\begin{eqnarray}
R_{sum}^A\hspace{-0.2cm}&\approx&\hspace{-0.2cm}\log_2
\left(1+\frac{M\rho_t}{K}\right)+(K-1)\log_2
\left(1+\frac{M\rho_t}{K}\left(1-\frac{2(K-1)}{c{M}}\right)\right).
\end{eqnarray}
It can be seen that the sum-rates of ICNS (the same as that of INS) and the ideal ZF have the similar increasing speed with respect to $M$, while the speed for MRT is smaller especially for more correlated channel (smaller $c$) and/or high transmission power. An example with $c=0.5$, $\rho_t=10$ and $K=10$ is given in Fig. \ref{fixed_K_finite or infiniely M} where both the sum-rates of ICNS and INS approach that of the ideal ZF for large $M$ while the sum-rate of MRT has much slower convergence rate.

\section{Numerical Results}\label{simulation}
In this section, simulation results are given to show the performance of the proposed schemes and its comparison with benchmarks, i.e., INS, DNS, TNS and CNS. The relaxation parameters in the proposed schemes and INS are all set according to \eqref{heuristic_omega}. Meanwhile, the analytical results in Theorem \ref{theorem 1} and \ref{theorem 2} will be verified. We consider two practical cases, i.e., 1) finite $K$ and growing large but finite $M$ and 2) growing $K$ and $M$ with fixed ratio.

\begin{figure}[t]
\centering
\includegraphics[scale=0.6]{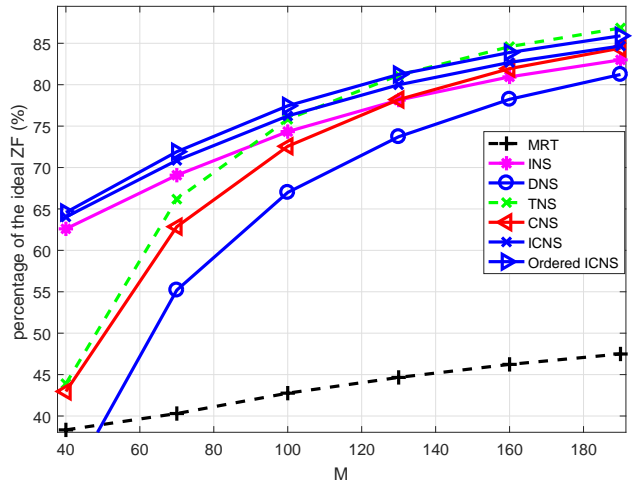}
\vspace{-7mm}
\caption{Comparison between the proposed schemes and existing ones for finite $K$ in terms of the ratio of their sum-rates to that of the ideal ZF. $c=0.5$, $\rho_t=10$, $K=10$. }\label{SIMU1}
\vspace{-10mm}
\end{figure}
For the case of fixed user number $K$ and increasing BS antenna number $M$, the ratios of the sum-rates of proposed and existing low-complexity schemes to that of the ideal ZF are shown in Fig. \ref{SIMU1} where the channel correlation level $c=0.5$, the transmission power $\rho_t=10$ and $K=10$.
It can be seen that 1) with the designed relaxation parameter in \eqref{heuristic_omega}, INS can outperform DNS and CNS for correlated channels and practical $M$ and the advantage becomes more significant as $M$ decreases. This is not explicitly shown in existing works. 2) The proposed schemes outperform all existing schemes except TNS. Compared with TNS which has higher complexity, the proposed schemes are largely better for small $M$, but TNS is slightly better in sum-rate than ICNS and the ordered ICNS for $M>100$ and $M>130$, respectively.
3) The sum-rate of the ordered ICNS is better than that of ICNS while the latter is better than that of INS. This validates the advantage resulted from the more careful handling of the user interference on the inversion approximation in ICNS as analytically proved in Section \ref{case2} and shows the benefit of further exploiting the multi-user diversity in the ordered ICNS.

\begin{figure*}[t]
  \normalsize
  \centering
  \hspace{-0.9cm}
  \begin{minipage}[t]{0.48\textwidth}
    \centering
    \includegraphics[scale=0.6]{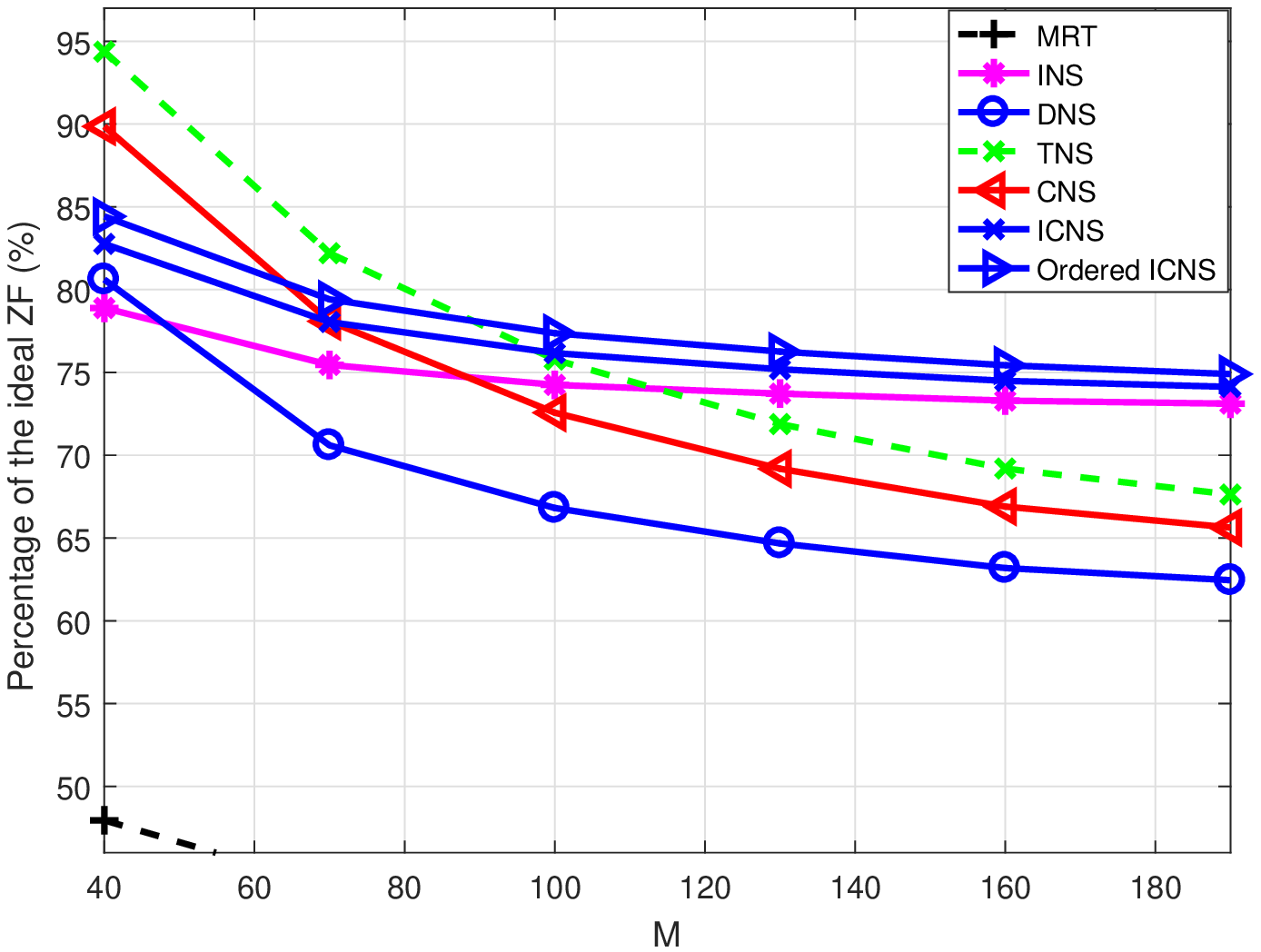}\vspace{-5mm}
    \caption{Comparison between proposed schemes and existing ones for increasing $K$ with $M$ in terms of the ratio of their sum-rates to that of the ideal ZF. $c=0.5$, $\rho_t=10$, $r=0.1$.}\label{SIMU2}
  \end{minipage}
  \hspace{0.35cm}
  \begin{minipage}[t]{0.48\textwidth}
    \centering
    \includegraphics[scale=0.6]{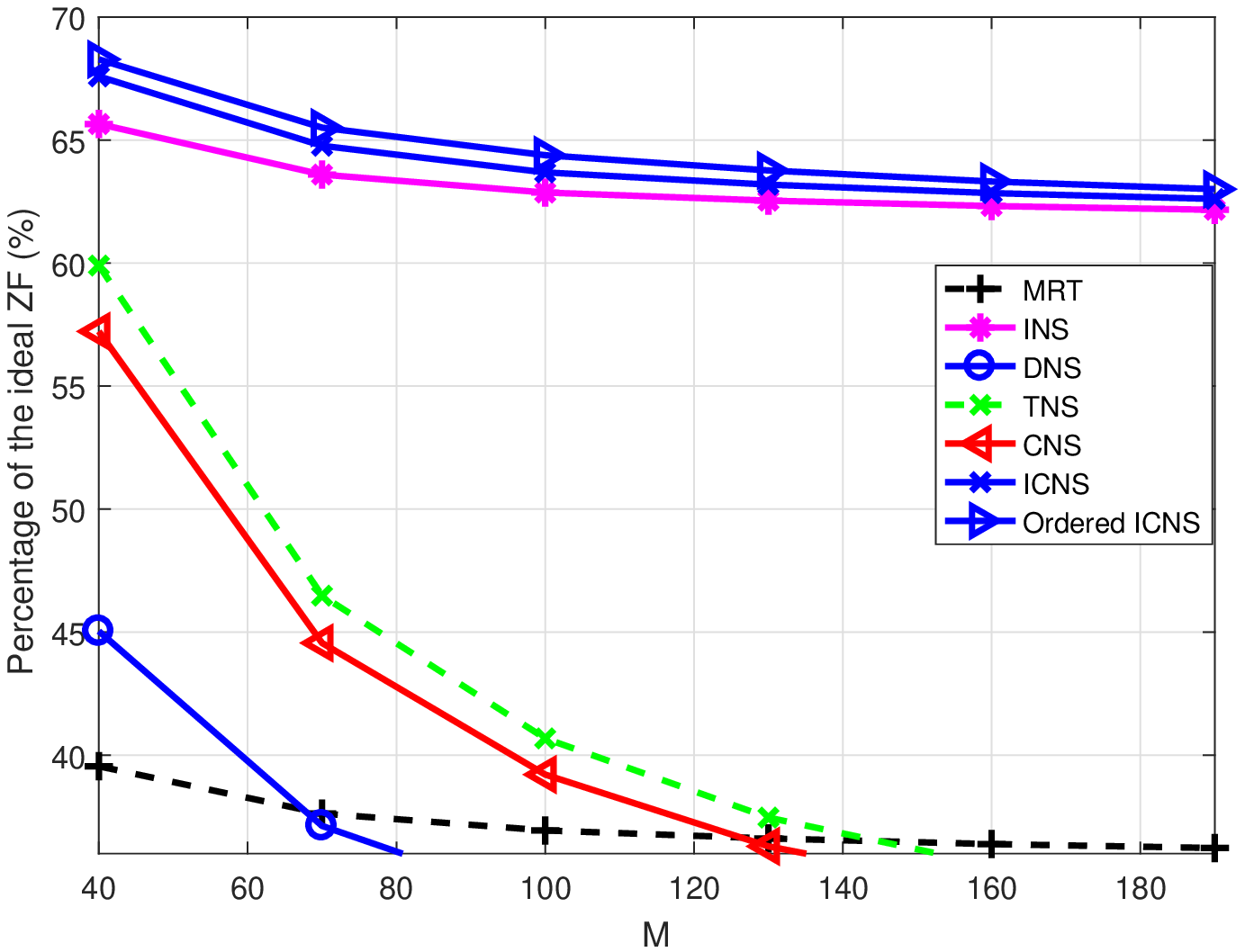}\vspace{-5mm} 
    \caption{Comparison between proposed schemes and existing ones for increasing $K$ with $M$ in terms of the ratio of their sum-rates to that of the ideal ZF. $c=0.5$, $\rho_t=10$, $r=0.2$.}\label{SIMU3}
  \end{minipage}
  \vspace{-7mm}
\end{figure*}
For the case of increasing $M$ and $K$ with fixed $r=K/M$, the ratios of the sum-rates of these low-complexity schemes to that of the ideal ZF are shown in Fig. \ref{SIMU2} and \ref{SIMU3} where $c=0.5$, $\rho_t=10$ and $r=0.1, 0.2$, respectively.
It can be seen that 1) the proposed ICNS and ordered ICNS schemes are superior over most existing ones
and the advantage becomes larger for larger $r$.
2) Further, both proposed schemes outperform INS, but the advantage becomes smaller for larger $r$ and larger $M$. This is because that the advantage of the proposed schemes over INS results from adding the non-diagonal elements of one column into the precondition matrix. Specifically, the number of interference terms that are considered in the proposed design is $K-1$ whose ratio to the whole number of interference terms $K^2-K$ becomes negligible when $K$ and $M$ increase. However, for practical range of $M$, this advantage for ICNS/ordered ICNS with affordable small extra complexity cost compared with INS is desirable.

\begin{figure*}[t]
  \normalsize
  \centering
  \hspace{-0.9cm}
  \begin{minipage}[t]{0.48\textwidth}
    \centering
    \includegraphics[scale=0.6]{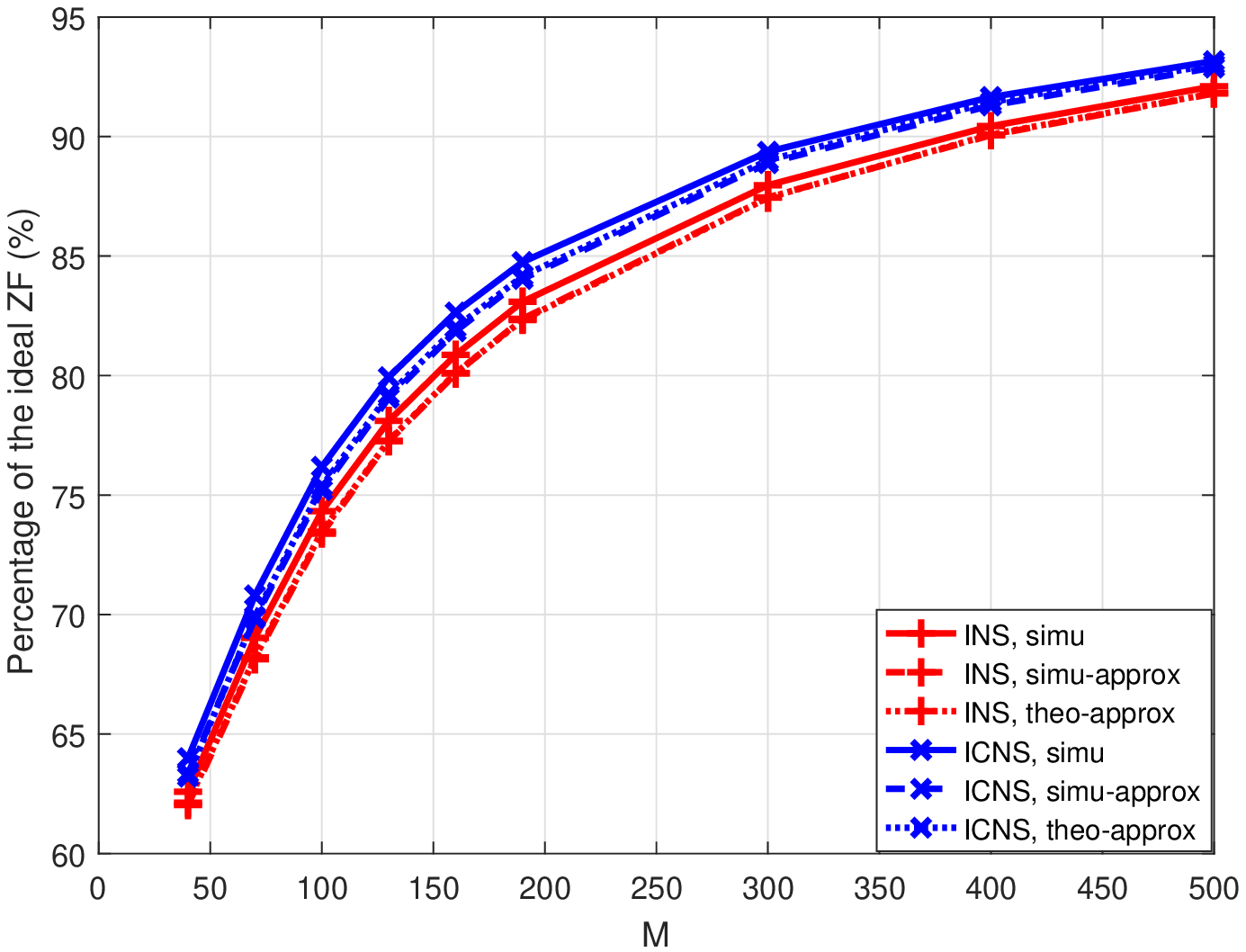}\vspace{-7mm}
    \caption{Validation of sum-rate approximations of INS and ICNS in terms of their ratio to the sum-rate of ideal ZF for finite $K$. $c=0.5$, $\rho_t=10$, $K=10$.}\label{K10_SNR10_analytical}
  \end{minipage}
  \hspace{0.385cm}
  \begin{minipage}[t]{0.48\textwidth}
    \centering
    \includegraphics[scale=0.6]{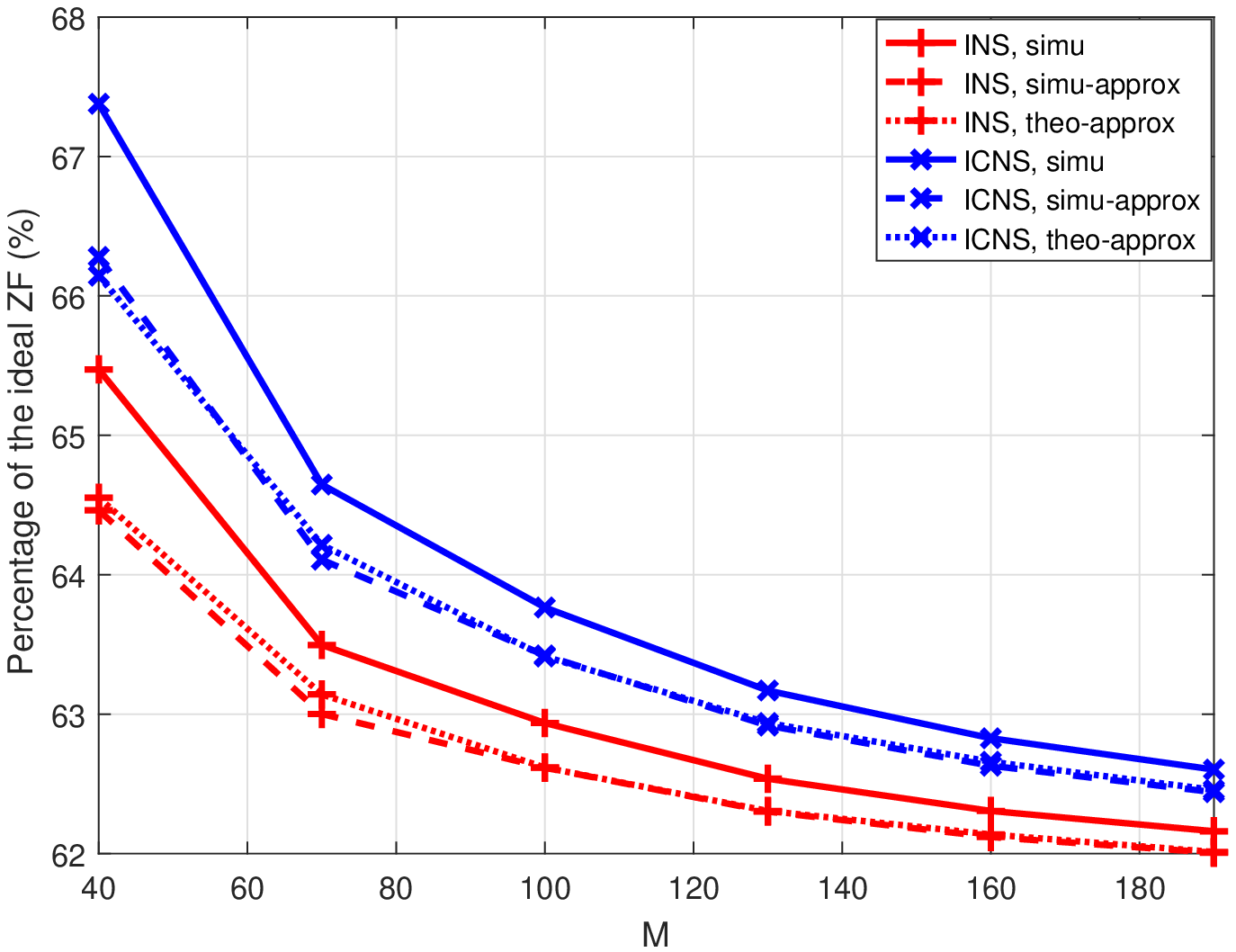}\vspace{-7mm} 
    \caption{Validation of sum-rate approximations of INS and ICNS in terms of their ratio to the sum-rate of ideal ZF for increasing $K$ with $M$. $c=0.5$, $\rho_t=10$, $r=0.2$.}\label{r0_4_SNR10_analytical}
  \end{minipage}
\vspace{-0.7cm}
\end{figure*}

For the verification of the analytical results in this paper, due to space limit, we only consider the closed-form sum-rate approximations given in Theorem \ref{theorem 1} and \ref{theorem 2} in Fig. \ref{K10_SNR10_analytical} and Fig. \ref{r0_4_SNR10_analytical}, respectively.
It can be seen that for both the case of fixed $K$ and growing large $M$ and the case of increasing $K$ and $M$ with fixed ratio $r$, the derived closed-form sum-rate approximations (denoted as \textit{theo-approx}) well match the simulated approximations as given in \eqref{sum-rate_a_INS_def} (denoted as \textit{simu-approx}). Meanwhile, the gap between the simulated sum-rates (\textit{simu}) and the approximations itself is small and decreases as $M$ grows, which shows the effectiveness of the derived results and corresponding comparison analysis.

\section{Conclusion}
For massive MIMO downlink, we studied the first-order NS expansion based low-complexity approximate ZF prcoding. Different from existing NS based schemes, for the proposed ICNS scheme, an effective relaxation parameter and one user's channel interference to others are jointly introduced into the construction of its precondition matrix. The proposed ordered ICNS further exploits the multi-user diversity gain based on ICNS. To study the performance loss of ICNS due to the matrix inversion approximation compared with the ideal ZF and its performance gain over the competitive benchmark INS,  closed-form sum-rate approximations of ICNS and INS were derived based on which explicit analysis for three typical massive MIMO scenarios were provided.
Finally, simulations verify our analytical results and the better performance-complexity tradeoff of the proposed schemes over the ideal ZF, INS and other existing low-complexity ZF precodings
for massive MIMO systems with correlated channels, practical large number of antennas, and not-so-small loading factor.


%

\appendices

{\centerline{\textbf{Appendix A: The Proof of Lemma \ref{Lemma 1}}}}
For the massive MIMO channel with correlation $c$, from \eqref{channel-model}, we have
\begin{equation}
{\bf H}={\bf R}^{1/2}{\bf Z},
\end{equation}
where ${\bf Z}=[{\bf z}_1,...,{\bf z}_K]$ whose columns are independent from each other. Then
\begin{equation}\label{eq63}
{\bf G}=\frac{{\bf Z}^H{\bf R}^{1/2}{\bf R}^{1/2}{\bf Z}}{M}= \frac{{\bf \tilde Z}^H{\bf \tilde Z}}{c{M}},
\end{equation}
where the second equality follows from \eqref{correlation model} and the definition of ${\bf \tilde Z}={\bf A}^H{\bf Z}$. Notice that $\tilde{\bf Z}$ is a $cM\times K$ matrix, which is different to $\bf H$.
Since the $k$-th column of ${\bf \tilde Z}$ satisfies ${\bf \tilde z}_k \sim \mathcal{CN}({\bf 0},{\bf I}_{c M})$, with the help of the Marchenko-Pastur distribution \cite{Prabhu_Approx}, the asymptotic maximum and minimum eigenvalue of $\bf { G}$ can be expressed respectively as $\bar{a}$ and $\bar{b}$ in \eqref{bar_a_b}.
From \eqref{optimal-omega-iid}, the relaxation parameter for INS that maximizes the asymptotic
convergence speed of the NS is as in \eqref{heuristic_omega}.

{\centerline{\textbf{Appendix B: The Proof of Theorem \ref{theorem 1}}}}

First we give Lemma \ref{lemma 2} as the preliminary for the subsequent derivations. Recall that
$\tilde{\bf Z}={\bf A}^H {\bf Z} $ and
${\bf{\tilde z}}_i$ is the $i$-th column of ${\bf{\tilde Z}}$ in \eqref{eq63}.
\begin{lemma}\label{lemma 2}
For all $i \ne j$,
\begin{eqnarray}
\label{eq38}\hspace{-1.3cm}&&{\rm E}\left\{ {{{\left| {{\bf{\tilde z}}_i^H{{\bf{\tilde z}}_j}} \right|}^2}} \right\} \hspace{-0.1cm}=\hspace{-0.1cm} c{M}, \quad {\rm E}\left\{ {{{\left| {{\bf{\tilde z}}_i^H{{\bf{\tilde z}}_j}} \right|}^2}{\bf{\tilde z}}_i^H{{\bf{\tilde z}}_j}} \right\}
\underset{a.s.}{\overset{M\rightarrow\infty}{\longrightarrow}}
0,\\
\label{eq39-1}\hspace{-1.3cm}&&{\rm E}\left\{\hspace{-0.1cm} {{{\left| {{\bf{\tilde z}}_i^H{{\bf{\tilde z}}_j}} \right|}^4}} \hspace{-0.05cm}\right\}\hspace{-0.1cm}/M^2 \hspace{0.1cm} \hspace{-0.1cm}\underset{a.s.}{\overset{M\rightarrow\infty}{\longrightarrow}}\hspace{-0.1cm} 2{c^2 }, \quad  {\rm E}\left\{ {{{\left| {{\bf{\tilde z}}_i^H{{\bf{\tilde z}}_j}} \right|}^6}} \right\}\hspace{-0.1cm}/{ M}^3 \underset{a.s.}{\overset{M\rightarrow\infty}{\longrightarrow}} 6c^3{},
\end{eqnarray}
where $a.s.$ denotes the almost sure convergence.
For all $ i$,
\begin{eqnarray}\label{pre_i_i}
\hspace{-1.9cm}&&{\rm E}\left\{ {{{\left| {{\bf{\tilde z}}_i^H{{\bf{\tilde z}}_i}} \right|}^2}} \right\} \hspace{-0.1cm}=\hspace{-0.1cm} c^2{M}^2+ c{M},\\
 \hspace{-1.9cm}&&{\rm E}\left\{ {{{\left| {{\bf{\tilde z}}_i^H{{\bf{\tilde z}}_i}} \right|}^3}} \right\} \hspace{-0.1cm}=\hspace{-0.1cm} c^3{M}^3+ 3c^2{M}^2+2c{M},\\
\label{eq42}\hspace{-1.9cm}&&{\rm E}\left\{ {{{\left| {{\bf{\tilde z}}_i^H{{\bf{\tilde z}}_i}} \right|}^4}} \right\} \hspace{-0.1cm}=\hspace{-0.1cm} c^4{M}^4+ 6c^3{M}^3+11c^2{M}^2+6c{M},\\ \label{eq68}
\label{eq43}\hspace{-1.9cm}&&{\rm E}\left\{ {{{\bf{\tilde z}}_i}{\bf{\tilde z}}_i^H{{\bf{\tilde z}}_i}{\bf{\tilde z}}_i^H} \right\} \hspace{-0.1cm}=\hspace{-0.1cm} \left( {c{M} + 1} \right){{\bf{I}}_{c{M}}}, \quad {\rm E}\left\{ {{{\bf{\tilde z}}_i}{\bf{\tilde z}}_i^H{{\bf{\tilde z}}_i}} \right\} = {\bf 0}.
\end{eqnarray}
\end{lemma}
\begin{IEEEproof}
Based on the central limit theorem \cite{Cheng_Performance}, ${\bf{\tilde z}}_i^H{{\bf{\tilde z}}_j}/{{\sqrt {c M} }} \sim \mathcal{CN}\left( {0,1} \right)$ when $M\rightarrow \infty$. Thus the second formula in \eqref{eq38} and \eqref{eq39-1} are derived with the help of the moments of normal distribution.
Since ${\bf{\tilde z}}_i^H{{\bf{\tilde z}}_i}$ follows the Gamma distribution with shape $c{M}$ and scale $1$, $|{\bf{\tilde z}}_i^H{{\bf{\tilde z}}_i}|^n,n=2,3,4$ follow the generalized gamma distribution and \eqref{pre_i_i}-\eqref{eq42} are obtained via calculating their expectations. For the first equation in \eqref{eq43},
since the $(l,k)$-th element of ${{{\bf{\tilde z}}_i}{\bf{\tilde z}}_i^H{{\bf{\tilde z}}_i}{\bf{\tilde z}}_i^H}$  is $\sum_{n=1}^{cM}{\tilde z}_{i,l}{\tilde z}^*_{i,n}{\tilde z}_{i,n}{\tilde z}^*_{i,k}$, its expectation is calculated with the help of the moments of normal distribution. The derivation for the last equation in \eqref{eq43} is similar.
\end{IEEEproof}

Define ${\bf F}_1=\frac{2}{\omega}{\bf G}-\frac{1}{\omega^2}{\bf G}^2$, we have
\begin{eqnarray}\label{eq71}
{\left[ {{{\bf{F}}_1}} \right]_{kj}} \hspace{-0.2cm}&=&\hspace{-0.2cm} \frac{{\rm{2}}}{\omega }{\left[ {{\bf{ G}}} \right]_{{{kj}}}} - \frac{{\rm{1}}}{{{\omega ^2}}}\sum\limits_{n = 1}^K {{{\left[ {{\bf{ G}}} \right]}_{{{kn}}}}{{\left[ {{\bf{ G}}} \right]}_{{{nj}}}}}, \forall k,j.
\end{eqnarray}
Recall that ${\left[ {{\bf{ G}}} \right]_{{{kj}}}} = {{\bf{\tilde z}}_k^H{{{\bf{\tilde z}}}_j}}/{{(c M)}}, \forall k,j$ (refer to \eqref{eq63}), thus
\begin{eqnarray}
\nonumber{\rm E}\left\{ {{{\left| {{{\left[ {{{\bf{F}}_1}} \right]}_{kk}}} \right|}^2}} \right\} \hspace{-0.1cm}&=&\hspace{-0.1cm} {\rm E}\left\{ {{{\left| {{{\left[ {{{\bf{F}}_1}} \right]}_{11}}} \right|}^2}} \right\},\forall k = 2,...,K,\\
{\rm E}\left\{ {{{\left| {{{\left[ {{{\bf{F}}_1}} \right]}_{kj}}} \right|}^2}} \right\} \hspace{-0.1cm}&=&\hspace{-0.1cm} {\rm E}\left\{ {{{\left| {{{\left[ {{{\bf{F}}_1}} \right]}_{12}}} \right|}^2}} \right\},\forall k \ne j.
\end{eqnarray}
Based on Lemma \ref{lemma 2}, ${\rm E}\left\{ {{{\left| {{{\left[ {{{\bf{F}}_1}} \right]}_{11}}} \right|}^2}} \right\}$ and ${\rm E}\left\{ {{{\left| {{{\left[ {{{\bf{F}}_1}} \right]}_{12}}} \right|}^2}} \right\}$ can be derived via some tedious calculations.

For $\beta_I$, from \eqref{eq3_1} and \eqref{W-INS} we have
\begin{eqnarray}\label{eq73}
\nonumber{\rm E}\{{\rm tr}\{{{\bf{W}}_I}{\bf{W}}_I^H\}\} \hspace{-0.1cm}&=&\hspace{-0.1cm} {\rm E}\left\{ {{\rm tr}\left\{ {\frac{{{\beta _I}}}{M}{\bf{H}}\left( {\frac{2}{\omega }{{\bf{I}}_K} - \frac{1}{{{\omega ^2}}}{\bf{G}}} \right)\frac{{{\beta _I}}}{M}{{\left( {\frac{2}{\omega }{{\bf{I}}_K} - \frac{1}{{{\omega ^2}}}{\bf{G}}} \right)}^H}{{\bf{H}}^H}} \right\}} \right\}\\
\nonumber\hspace{-0.1cm}&=&\hspace{-0.1cm} \frac{{\beta _I^2}}{M}{\rm E}\left\{ {{\rm tr}\left\{ {\left( {\frac{2}{\omega }{\bf{G}} - \frac{1}{{{\omega ^2}}}{{\bf{G}}^2}} \right){{\left( {\frac{2}{\omega }{{\bf{I}}_K} - \frac{1}{{{\omega ^2}}}{\bf{G}}} \right)}^H}} \right\}} \right\} \\
\hspace{-0.1cm}&=&\hspace{-0.1cm} \frac{{\beta _I^2}}{M}{\rm E}\left\{ {{\rm tr}\left\{ {{{\bf{F}}_1}{{\left( {\frac{2}{\omega }{{\bf{I}}_K} - \frac{1}{{{\omega ^2}}}{\bf{ G}}} \right)}^H}} \right\}} \right\}=1.
\end{eqnarray}
Then, from \eqref{eq71} we have
\begin{eqnarray}
\nonumber\hspace{-0.2cm}{\left[ {{{\bf{F}}_1}{{\left( {\frac{2}{\omega }{{\bf{I}}_K} - \frac{1}{{{\omega ^2}}}{\bf{ G}}} \right)}^H}} \right]_{kk}} \hspace{-0.5cm}\hspace{-0.2cm}&&= \left( {\frac{{\rm{2}}}{\omega }{{\left[ {\bf{ G}} \right]}_{{{kk}}}} - \frac{{\rm{1}}}{{{\omega ^2}}}\sum\limits_{n = 1}^K {{{\left[ {\bf{ G}} \right]}_{{{kn}}}}{{\left[ {\bf{ G}} \right]}_{{{nk}}}}} } \right){\left( {\frac{{\rm{2}}}{\omega } - \frac{{\rm{1}}}{{{\omega ^2}}}{{\left[ {\bf{ G}} \right]}_{{{kk}}}}} \right)}\\
 &&+\hspace{-0.2cm} \sum\limits_{j = 1,j \ne k}^K {\left( {\frac{{\rm{2}}}{\omega }{{\left[ {\bf{ G}} \right]}_{{{kj}}}} - \frac{{\rm{1}}}{{{\omega ^2}}}\sum\limits_{n = 1}^K {{{\left[ {\bf{ G}} \right]}_{{{kn}}}}{{\left[ {\bf{ G}} \right]}_{{{nj}}}}} } \right){{\left( { - \frac{{\rm{1}}}{{{\omega ^2}}}{{\left[ {\bf{ G}} \right]}_{{{jk}}}}} \right)}}}, \quad \forall k.
\end{eqnarray}
Since
\begin{eqnarray}
{\rm E}\left\{{\left[ {{{\bf{F}}_1}{{\left( {\frac{2}{\omega }{{\bf{I}}_K} - \frac{1}{{{\omega ^2}}}{\bf{ G}}} \right)}^H}} \right]_{kk}}\right\}&=&{\rm E}\left\{{\left[ {{{\bf{F}}_1}{{\left( {\frac{2}{\omega }{{\bf{I}}_K} - \frac{1}{{{\omega ^2}}}{\bf{ G}}} \right)}^H}} \right]_{jj}}\right\}, \quad \text{for}\quad k\ne j,\end{eqnarray}
we have
\begin{eqnarray}
{\rm E}\left\{ {{\rm tr}\left\{ {{{\bf{F}}_1}{{\left( {\frac{2}{\omega }{{\bf{I}}_K} - \frac{1}{{{\omega ^2}}}{\bf{G}}} \right)}^H}} \right\}} \right\}&=&K {\rm E}\left\{{\left[ {{{\bf{F}}_1}{{\left( {\frac{2}{\omega }{{\bf{I}}_K} - \frac{1}{{{\omega ^2}}}{\bf{G}}} \right)}^H}} \right]_{11}}\right\}
\end{eqnarray}
which can be obtained via some tedious calculations based on Lemma \ref{lemma 2}. Finally, from \eqref{eq73}, $\beta_I$ can be obtained. Based on the results of ${\rm E}\left\{ {{{\left| {{{\left[ {{{\bf{F}}_1}} \right]}_{11}}} \right|}^2}} \right\}$ and ${\rm E}\left\{ {{{\left| {{{\left[ {{{\bf{F}}_1}} \right]}_{12}}} \right|}^2}} \right\}$ and $\beta_I$, \eqref{eq39}-\eqref{eq40} can be obtained via eliminating the common factor ${1}/{\omega^2}$ with the help of \eqref{H_BETA} and \eqref{sum-rate_a_INS_def}.

{\centerline{\textbf{Appendix C: The Proof of Theorem \ref{theorem 2}}}}

Define ${{\bf{F}}_2} = 2{\bf{ G D}}_A^{ - 1} - {\left( {{\bf{ G D}}_A^{ - 1}} \right)^2}$. From \eqref{inverse-precond} we have
\begin{eqnarray}\label{eq79}
\nonumber&&{\left[ {{{\bf{F}}_2}} \right]_{11}} = \frac{{2}}{\omega }\left( {{{\left[ {{\bf{ G}}} \right]}_{{\rm{11}}}} - \frac{1}{\omega }\sum\limits_{k = 2}^K {{{\left| {{{\left[ {{\bf{ G}}} \right]}_{{{1k}}}}} \right|}^2}} } \right) \\
&&- \frac{{\rm{1}}}{{{\omega ^2}}}\left( {{{\left( {{{\left[ {{\bf{ G}}} \right]}_{{\rm{11}}}} - \frac{1}{\omega }\sum\limits_{k = 2}^K {{{\left| {{{\left[ {{\bf{ G}}} \right]}_{{{1k}}}}} \right|}^2}} } \right)}^2} + \sum\limits_{n = 2}^K {{{\left[ {{\bf{ G}}} \right]}_{{{1n}}}}\left( {{{\left[ {{\bf{ G}}} \right]}_{{{n1}}}} - \frac{1}{\omega }\sum\limits_{k = 2}^K {{{\left[ {{\bf{ G}}} \right]}_{{{nk}}}}{{\left[ {{\bf{ G}}} \right]}_{{{k1}}}}} } \right)} } \right),
\end{eqnarray}
\begin{eqnarray}\label{eq80}
\hspace{0.4cm}{\left[ {{{\bf{F}}_2}} \right]_{1j}} = \frac{2}{\omega }{\left[ {{\bf{ G}}} \right]_{{{1j}}}} - \frac{{\rm{1}}}{{{\omega ^2}}}\left( { - \frac{1}{\omega }\sum\limits_{k = 2}^K {{{\left| {{{\left[ {{\bf{ G}}} \right]}_{{{1k}}}}} \right|}^2}{{\left[ {{\bf{ G}}} \right]}_{{{1j}}}}}  + \sum\limits_{k = 1}^K {{{\left[ {{\bf{ G}}} \right]}_{{{1k}}}}{{\left[ {{\bf{ G}}} \right]}_{{{kj}}}}} } \right), \forall j\ge2,
\end{eqnarray}
\begin{eqnarray}\label{eq81}
\nonumber&&\hspace{-0.8cm}{\left[ {{{\bf{F}}_2}} \right]_{j1}} = 2\frac{1}{\omega }\left( {{{\left[ {{\bf{ G}}} \right]}_{{{j1}}}} - \frac{1}{\omega }\sum\limits_{k = 2}^K {{{\left[ {{\bf{ G}}} \right]}_{{{jk}}}}{{\left[ {{\bf{ G}}} \right]}_{{{k1}}}}} } \right)\\
&&\hspace{-0.8cm}- \frac{{\rm{1}}}{{{\omega ^2}}}\left( \begin{array}{l}
\left( {{{\left[ {{\bf{ G}}} \right]}_{{{j1}}}} - \frac{1}{\omega }\sum\limits_{k = 2}^K {{{\left[ {{\bf{ G}}} \right]}_{{{jk}}}}{{\left[ {{\bf{ G}}} \right]}_{{{k1}}}}} } \right)\left( {{{\left[ {{\bf{ G}}} \right]}_{{\rm{11}}}} - \frac{1}{\omega }\sum\limits_{k = 2}^K {{{\left| {{{\left[ {{\bf{ G}}} \right]}_{{{1k}}}}} \right|}^2}} } \right)\\
 + \sum\limits_{n = 2}^K {{{\left[ {{\bf{ G}}} \right]}_{{{jn}}}}\left( {{{\left[ {{\bf{ G}}} \right]}_{{{n1}}}} - \frac{1}{\omega }\sum\limits_{k = 2}^K {{{\left[ {{\bf{ G}}} \right]}_{{{nk}}}}{{\left[ {{\bf{ G}}} \right]}_{{{k1}}}}} } \right)}
\end{array} \right), \forall j\ge2,
\end{eqnarray}
\begin{eqnarray} \label{eq82}
{\left[ {{{\bf{F}}_2}} \right]_{jj}} = \frac{{\rm{2}}}{\omega }{\left[ {{\bf{ G}}} \right]_{jj}} - \frac{{\rm{1}}}{{{\omega ^2}}}\left( {\left( {{{\left[ {{\bf{ G}}} \right]}_{{{j1}}}} - \frac{1}{\omega }\sum\limits_{k = 2}^K {{{\left[ {{\bf{ G}}} \right]}_{{{jk}}}}{{\left[ {{\bf{ G}}} \right]}_{{{k1}}}}} } \right){{\left[ {{\bf{ G}}} \right]}_{{{1j}}}} + \sum\limits_{k = 2}^K {{{\left[ {{\bf{ G}}} \right]}_{{{jk}}}}{{\left[ {{\bf{ G}}} \right]}_{{{kj}}}}} } \right), \forall j\ge2,
\end{eqnarray}
and
\begin{eqnarray} \label{eq83}
 \nonumber\hspace{-0.9cm}{\left[ {{{\bf{F}}_2}} \right]_{jm}} = \frac{{2}}{\omega }{\left[ {{\bf{ G}}} \right]_{{{jm}}}} - \frac{{\rm{1}}}{{{\omega ^2}}}\left( {\left( {{{\left[ {{\bf{ G}}} \right]}_{{{j1}}}} - \frac{1}{\omega }\sum\limits_{k = 2}^K {{{\left[ {{\bf{ G}}} \right]}_{{{jk}}}}{{\left[ {{\bf{ G}}} \right]}_{{{k1}}}}} } \right){{\left[ {{\bf{ G}}} \right]}_{{{1m}}}}{\rm{ + }}\sum\limits_{k = 2}^K {{{\left[ {{\bf{ G}}} \right]}_{{{jk}}}}{{\left[ {{\bf{ G}}} \right]}_{{{km}}}}} } \right),\\
 \forall j\ge2,m\ge2,m\ne j.
\end{eqnarray}
It can be seen from \eqref{eq79}-\eqref{eq83} that
\begin{eqnarray}
{\rm E}\left\{ {{{\left| {{{\left[ {{{\bf{F}}_2}} \right]}_{1k}}} \right|}^2}} \right\} \hspace{-0.1cm}&=&\hspace{-0.1cm} {\rm E}\left\{ {{{\left| {{{\left[ {{{\bf{F}}_2}} \right]}_{12}}} \right|}^2}} \right\},\forall k \ge 2, \\
{\rm E}\left\{ {{{\left| {{{\left[ {{{\bf{F}}_2}} \right]}_{k1}}} \right|}^2}} \right\} \hspace{-0.1cm}&=&\hspace{-0.1cm} {\rm E}\left\{ {{{\left| {{{\left[ {{{\bf{F}}_2}} \right]}_{21}}} \right|}^2}} \right\},\forall k \ge 2,\\
{\rm E}\left\{ {{{\left| {{{\left[ {{{\bf{F}}_2}} \right]}_{kk}}} \right|}^2}} \right\} \hspace{-0.1cm}&=&\hspace{-0.1cm} {\rm E}\left\{ {{{\left| {{{\left[ {{{\bf{F}}_2}} \right]}_{22}}} \right|}^2}} \right\},\forall k \ge 2,\\
{\rm E}\left\{ {{{\left| {{{\left[ {{{\bf{F}}_2}} \right]}_{km}}} \right|}^2}} \right\} \hspace{-0.1cm}&=&\hspace{-0.1cm} {\rm E}\left\{ {{{\left| {{{\left[ {{{\bf{F}}_2}} \right]}_{23}}} \right|}^2}} \right\},\forall k \ge 2,m \ge 2,k \ne m.
\end{eqnarray}
Thus the problem of calculating ${\rm E}\left\{ {{{\left| {{{\left[ {{{\bf{F}}_2}} \right]}_{km}}} \right|}^2}} \right\},\forall k,m$ can be transformed to that of ${\rm E}\left\{ {{{\left| {{{\left[ {{{\bf{F}}_2}} \right]}_{11}}} \right|}^2}} \right\}$, ${\rm E}\left\{ {{{\left| {{{\left[ {{{\bf{F}}_2}} \right]}_{12}}} \right|}^2}} \right\}$,  ${\rm E}\left\{ {{{\left| {{{\left[ {{{\bf{F}}_2}} \right]}_{22}}} \right|}^2}} \right\}$, ${\rm E}\left\{ {{{\left| {{{\left[ {{{\bf{F}}_2}} \right]}_{23}}} \right|}^2}} \right\}$ and ${\rm E}\left\{ {{{\left| {{{\left[ {{{\bf{F}}_2}} \right]}_{21}}} \right|}^2}} \right\}$. These terms can be derived via some tedious calculations based on Lemma \ref{lemma 2}, the observation that $cM\sum\nolimits_{k = 2}^K {{{\left| {{{\left[ {\bf{G}} \right]}_{1k}}} \right|}^2}}$ follows the Gamma distribution with shape $K-1$ and scale $1$, and three assumptions which are explained in the following.

Recall that $\left[ {\bf{G}} \right]_{ij}={\bf{\tilde z}}_i^H{{\bf{\tilde z}}_j}/{{{(c M)} }}$. Since ${\rm E}\{{\left[ {\bf{G}} \right]_{{{ii}}}}\}=1, \forall i$ and its variance is $1/(cM)$, i.e., the ratio of its variance to its mean square is negligible for large $M$, ${\left[ {\bf{G}} \right]_{{{ii}}}}$ can be approximated as a deterministic value \cite{Cheng_Performance}, i.e.,
\begin{eqnarray}
\label{ass1}{\left[ {\bf{G}} \right]_{{{ii}}}} \approx 1, \forall i, \quad \text{when} \quad M\gg 1.
\end{eqnarray}
Meanwhile, for $K\gg1$, we have
\begin{eqnarray}
\label{ass2} {\left| {{{\left[ {\bf{G}} \right]}_{{{12}}}}} \right|^2} + \sum\nolimits_{k = 3}^K {{{\left| {{{\left[ {\bf{G}} \right]}_{{{1k}}}}} \right|}^2}}  \approx \sum\nolimits_{k = 3}^K {{{\left| {{{\left[ {\bf{G}} \right]}_{{{1k}}}}} \right|}^2}}.
\end{eqnarray}
which follows from 1) the means of the left and right sides of the equation differ by a factor of ${(K-1)}/{(K-2)}$; 2) the ratio of their variances to their mean squares decrease linearly with $K$ (i.e., approximately deterministic for $K\gg 1$).
These two assumptions are used in the calculations of ${\rm E}\{ {{{\left| {{{\left[ {{{\bf{F}}_2}} \right]}_{21}}} \right|}^2}} \}$. For the calculation of ${\rm E}\{ {{{\left| {{{\left[ {{{\bf{F}}_2}} \right]}_{11}}} \right|}^2}} \}$, besides the assumption in \eqref{ass1}, another used assumption is
\begin{equation}\label{ass3}
\sum\limits_{n = 2}^K {\sum\limits_{k = 2,k \ne n}^K {{{\left[ {\bf{G}} \right]}_{n1}}{{\left[ {\bf{G}} \right]}_{1k}}{{\left[ {\bf{G}} \right]}_{kn}}} }  \approx \frac{{K - 2}}{{cM}}\sum\limits_{k = 2}^K {{{\left| {{{\left[ {\bf{G}} \right]}_{1k}}} \right|}^2}},
\end{equation}
which follows from
\begin{eqnarray}\nonumber
\sum\limits_{k = 2,k \ne n}^K {{{\left[ {\bf{G}} \right]}_{1k}}{{\left[ {\bf{G}} \right]}_{kn}}} \hspace{-0.2cm}& = &\hspace{-0.2cm} {\bf{\tilde z}}_1^H\left( {\sum\limits_{k = 2,k \ne n}^K {\frac{{{{{\bf{\tilde z}}}_k}{\bf{\tilde z}}_k^H}}{{{c^2}{M^2}}}} } \right){{\bf{\tilde z}}_n} = {\bf{\tilde z}}_1^H\frac{{K - 2}}{{{c^2}{M^2}}}\frac{1}{{K - 2}}\left( {\sum\limits_{k = 2,k \ne n}^K {{{{\bf{\tilde z}}}_k}{\bf{\tilde z}}_k^H} } \right){{\bf{\tilde z}}_n}\\
 \hspace{-0.2cm}&\approx&\hspace{-0.2cm} \frac{{K - 2}}{{{c^2}{M^2}}}{\bf{\tilde z}}_1^H{{\bf{\tilde z}}_n} = \frac{{K - 2}}{{cM}}{\left[ {\bf{G}} \right]_{1n}}, \quad \text{when} \quad K\gg 1.
\end{eqnarray}

For $\beta_A$, from \eqref{eq3_1} and \eqref{precoding} we have
\begin{eqnarray}
\nonumber\hspace{-0.5cm}{\rm E}\left\{ {{\rm tr}\left\{ {{{\bf{W}}_A}{\bf{W}}_A^H} \right\}} \right\}\hspace{-0.4cm}&{\rm{ = }}&\hspace{-0.4cm}{\rm E}\left\{ {{\rm tr}\left\{ {\frac{{{\beta _A}}}{M}{\bf{H}}\left( {2{\bf{D}}_A^{ - 1} - {\bf{D}}_A^{ - 1}{\bf{GD}}_A^{ - 1}} \right){{\left( {\frac{{{\beta _A}}}{M}{\bf{H}}\left( {2{\bf{D}}_A^{ - 1} - {\bf{D}}_A^{ - 1}{\bf{GD}}_A^{ - 1}} \right)} \right)}^H}} \right\}} \right\}\\ \nonumber
&{\rm{ = }}&\frac{{\beta _A^2}}{M}{\rm E}\left\{ {{\rm tr}\left\{ {\left( {2{\bf{GD}}_A^{ - 1} - {\bf{GD}}_A^{ - 1}{\bf{GD}}_A^{ - 1}} \right){{\left( {2{\bf{D}}_A^{ - 1} - {\bf{D}}_A^{ - 1}{\bf{GD}}_A^{ - 1}} \right)}^H}} \right\}} \right\}\\
&{\rm{ = }}&\frac{{\beta _A^2}}{M}{\rm E}\left\{ {{\rm tr}\left\{ {{{\bf{F}}_2}{{\left( {2{\bf{ D}}_A^{ - 1} - {\bf{ D}}_A^{ - 1}{\bf{ G D}}_A^{ - 1}} \right)}^H}} \right\}} \right\}.
\end{eqnarray}
Further, from
\eqref{inverse-precond} we have
\begin{equation}
{{\bf{F}}_2}{\left( {2{\bf{D}}_A^{ - 1} - {\bf{D}}_A^{ - 1}{\bf{G D}}_A^{ - 1}} \right)^H}{\rm{ = }}\frac{2}{\omega }{\bf{A}} - \frac{1}{{{\omega ^2}}}{\bf{B}},
\end{equation}
where
\begin{equation}
{\left[ {\bf{A}} \right]_{11}} = {\left[ {{{\bf{F}}_2}} \right]_{11}}, \quad {\left[ {\bf{A}} \right]_{kk}} = {\left[ {{{\bf{F}}_2}} \right]_{kk}} - \frac{{\left[ {{\bf{ G}}} \right]_{{{1k}}}}}{\omega }{\left[ {{{\bf{F}}_2}} \right]_{k1}}, \quad k\ge2,
\end{equation}
\begin{equation}
{\left[ {\bf{B}} \right]_{11}} = {\left[ {{{\bf{F}}_2}} \right]_{11}}\left( {\left[ {{\bf{ G}}} \right]_{{\rm{11}}} - \sum\limits_{k = 2}^K {\left[ {{\bf{ G}}} \right]_{{{k1}}}\frac{{\left[ {{\bf{ G}}} \right]_{{{1k}}}}}{\omega }} } \right) + \sum\limits_{k = 2}^K {{{\left[ {{{\bf{F}}_2}} \right]}_{1k}}\left[ {{\bf{ G}}} \right]_{{{k1}}}},
\end{equation}
and
\begin{eqnarray}
\nonumber{\left[ {\bf{B}} \right]_{kk}} \hspace{-0.2cm}&=&\hspace{-0.2cm} {\left[ {{{\bf{F}}_2}} \right]_{k1}}\left( { - \frac{{\left[ {{\bf{ G}}} \right]_{{{1k}}}}}{\omega }\left[ {{\bf{ G}}} \right]_{{\rm{11}}} + \left[ {{\bf{ G}}} \right]_{{{1k}}} - \sum\limits_{n = 2}^K {\left( { - \frac{{\left[ {{\bf{ G}}} \right]_{1k}}}{\omega }\left[ {{\bf{ G}}} \right]_{{{n1}}} + \left[ {{\bf{ G}}} \right]_{{{nk}}}} \right)\frac{{\left[ {{\bf{ G}}} \right]_{{{1n}}}}}{\omega }} } \right)\\
&+&\hspace{-0.2cm} \sum\limits_{n = 2}^K {{{\left[ {{{\bf{F}}_2}} \right]}_{kn}}\left( { - \frac{{\left[ {{\bf{ G}}} \right]_{{{1k}}}}}{\omega }\left[ {{\bf{ G}}} \right]_{{{n1}}} + \left[ {{\bf{ G}}} \right]_{{{nk}}}} \right)}, \quad k\ge2.
\end{eqnarray}
Therefore, the calculation of $\beta_A$ is transformed to the calculations of ${\rm E}\left\{ {{{\left[ {\bf{A}} \right]}_{11}}} \right\}$, ${\rm E}\left\{ {{{\left[ {\bf{A}} \right]}_{kk}}} \right\},k\ge 2$, ${\rm E}\left\{ {{{\left[ {\bf{B}} \right]}_{11}}} \right\}$ and ${\rm E}\left\{ {{{\left[ {\bf{B}} \right]}_{kk}}} \right\},k\ge 2$. Further, since for $k\ge 2$, $j\ge 2$ and $k\ne j$, ${\rm E}\left\{ {{{\left[ {\bf{A}} \right]}_{kk}}} \right\}={\rm E}\left\{ {{{\left[ {\bf{A}} \right]}_{jj}}} \right\}$ and ${\rm E}\left\{ {{{\left[ {\bf{B}} \right]}_{kk}}} \right\}={\rm E}\left\{ {{{\left[ {\bf{B}} \right]}_{jj}}} \right\}$, only ${\rm E}\left\{ {{{\left[ {\bf{A}} \right]}_{11}}} \right\}$, ${\rm E}\left\{ {{{\left[ {\bf{A}} \right]}_{22}}} \right\}$, ${\rm E}\left\{ {{{\left[ {\bf{B}} \right]}_{11}}} \right\}$ and ${\rm E}\left\{ {{{\left[ {\bf{B}} \right]}_{22}}} \right\}$ need to be calculated. These can be obtained via some tedious calculations based on Lemma \ref{lemma 2}.
Note that for ${\rm E}\left\{ {{{\left[ {\bf{B}} \right]}_{22}}} \right\}$, the approximations in \eqref{ass1} and \eqref{ass2} are also used to simplify the derivation procedure with negligible difference.

{\centerline{\textbf{Appendix D: The Proof of Corollary \ref{corollary1}}}}
Since the effective SINR for all users with INS are the same, the comparison of sum-rates can be transferred to the comparison of effective SINRs.
Recall that $\omega=1+{r}/c$. The effective SINR of INS in \eqref{sum-rate_INS} can be approximated as
\begin{equation}\label{SINR_INS_r}
\widetilde{\rm SINR}^{I}\approx\frac{{{\rho _t}}}{ r}\frac{1}{{ {1{\rm{ + }}\frac{{ r}c}{{{{\left( {{ r} + c} \right)}^2}}}} {\rm{ + }}{\rho _t}\frac{{ r}/c}{{{ r} + c}}}},
\end{equation}
where $\mathcal{O}(1/{M})$ terms are omitted in the numerator and denominator of $\widetilde{\rm SINR}^{I}$.
By replacing $K-1$ with $K$ in \eqref{sum-rate-MRT}, we have $\widetilde{{\rm{SINR}}}_{MRT} \approx \frac{{{\rho _t}}}{{ r}/c}\frac{1}{{{\rho _t} + c}}$ and
\begin{equation}
\left({ {1{\rm{ + }}\frac{{ r}c}{{{{\left( {{ r} + c} \right)}^2}}}} {\rm{ + }}{\rho _t}\frac{{ r}/c}{{{ r} + c}}}\right) - \frac{{\rho _t} + c}{c} = \frac{{\left( {c - {\rho _t}} \right){ r} - {\rho _tc}}}{{{{\left( {{ r} + c} \right)}^2}}}<0, \quad \text{if} \quad \rho_t>{{ r}c}/({ r}+c).
\end{equation}

The ratio of the effective SINR of INS to that of the ideal ZF in \eqref{sinr_ZF} can be written as
\begin{equation}\label{eq61}
Pr_{\rm INS}=\frac{\widetilde{\rm SINR}^{I}}{\rho_t\left(\frac{1}{{ r}}-\frac{1}{c}\right)}.
\end{equation}
From \eqref{SINR_INS_r}, \eqref{eq61} can be transformed to
\begin{equation}\label{eq65}
\frac{{ r}^3}{c^3}P{r_{{\rm{INS}}}}\left( { - 1 - \frac{{{\rho _t}}}{c}} \right) + \frac{{ r}^2}{c^2}\left( { - 2P{r_{{\rm{INS}}}} - 1} \right) + \frac{ r}{c}\left( {2P{r_{{\rm{INS}}}} + P{r_{{\rm{INS}}}}\frac{{{\rho _t}}}{c} - 2} \right) + P{r_{{\rm{INS}}}} - 1 = 0
\end{equation}
which is a standard cubic equation.
For $Pr_{\rm INS}=1$ and ${r}>0$, \eqref{eq65} becomes a quadratic equation, i.e.,
\begin{equation}
- \frac{{ r}^2}{c^2}\left( {1{\rm{ + }}\frac{{{\rho _t}}}{c}} \right) - 3\frac{ r}{c} + \frac{{{\rho _t}}}{c} = 0.
\end{equation}
It can be easily know that the only positive real root of this quadratic equation is
\begin{equation}
{{r}^*} =\frac{3-\sqrt{9+4\frac{\rho_t}{c}+4\frac{\rho_t^2}{c^2}}}{-2(1+\frac{\rho_t}{c})}c
\end{equation}
which can be simplified to \eqref{pos_root}.



\ifCLASSOPTIONcaptionsoff
  \newpage
\fi

\end{document}